%% file: main.tex
\documentclass[3p, final, times]{elsarticle}
\pdfoutput=1
\makeatletter
\def\ps@pprintTitle{%
 \let\@oddhead\@empty
 \let\@evenhead\@empty
 \def\@oddfoot{\centerline{\thepage}}%
 \let\@evenfoot\@oddfoot}
\makeatother
\journal{Information and Software Technology}
\usepackage[table,xcdraw]{xcolor}

\usepackage{upgreek}
\usepackage{amsmath,amssymb,amsfonts}
\usepackage{float}
\usepackage{threeparttable}
\usepackage{tabularx}
\usepackage{tcolorbox}
\usepackage{subcaption}
\usepackage{multirow}
\usepackage{longfbox}
\usepackage{makecell}
\usepackage{longtable}
\usepackage{enumitem}
\usepackage{textcomp}
\usepackage{hyperref}
\usepackage{booktabs}
\usepackage{natbib}
\usepackage{hyperref}
\hypersetup{
    colorlinks=true,
    linkcolor=blue,
    filecolor=magenta,      
    urlcolor=cyan,
}

\newcounter{finding_counter}
\newtoggle{comment}
\toggletrue{comment}

\newcommand{\rifat}[1] { 
    \iftoggle{comment}{
        \textcolor{blue}{\\ Rifat: #1\\}
     }
}
\newcommand{\ra}[1]{\textcolor{black}{#1}}

\newcommand{\boxtext}[1]{
    \begin{tcolorbox}
        \textbf{Finding \arabic{finding_counter}:} #1 \stepcounter{finding_counter}
    \end{tcolorbox}
}

\title{How Do Developers Discuss and Support New Programming Languages in Technical Q\&A Site?\\ \large{An Empirical Study of Go, Swift, and Rust in Stack Overflow}}

\begin{document}

\begin{frontmatter}

\author{Partha Chakraborty$^a$, Rifat Shahriyar$^a$, Anindya Iqbal$^a$, and Gias Uddin$^b$}
\address{$^a$Bangladesh University of Engineering and Technology and $^b$University of Calgary}

\input{Sections/Abstract}

\begin{keyword}
Stack Overflow\sep Swift\sep Go\sep Rust \sep New Language \sep Evolution
\end{keyword}

\end{frontmatter}

\input{Sections/Introduction}
\input{Sections/Background}

\input{Sections/Dev_Discussions}

\input{Sections/Dev_Support}
\input{Sections/Implication}
\input{Sections/Validity}
\input{Sections/RelatedWorks}
\input{Sections/Conclusion}



\begin{small} 
\bibliographystyle{elsarticle-num}
\bibliography{main}   
\end{small}

\end{document}

%% file: Sections/Abstract.tex
\begin{abstract}
\noindent\textbf{Context:} New programming languages (e.g., Swift, Go, Rust, etc.) are being introduced to provide a better opportunity for the developers to make software development robust and easy. At the early stage, a programming language is likely to have resource constraints that encourage the developers to seek help frequently from experienced peers active in Question-Answering (QA) sites such as Stack Overflow (SO).
\newline
\textbf{Objective:} In this study, we have formally studied the discussions on three popular new languages introduced after the inception of SO (2008) and match those with the relevant activities in GitHub whenever appropriate. For that purpose, we have mined 4,17,82,536 questions and answers from SO and 7,846 issue information along with 6,60,965 repository information from Github. Initially, the development of new languages is relatively slow compared to mature languages (e.g., C, C++, Java). The expected outcome of this study is to reveal the difficulties and challenges faced by the developers working with these languages so that appropriate measures can be taken to expedite the generation of relevant resources.
\newline
\textbf{Method:} We have used the Latent Dirichlet Allocation (LDA) method on SO's questions and answers to identify different topics of new languages. We have extracted several features of the answer pattern of the new languages from SO (e.g., time to get an accepted answer, time to get an answer, etc.) to study their characteristics. These attributes were used to identify difficult topics. We explored the background of developers who are contributing to these languages. We have created a model by combining Stack Overflow data and issues, repository, user data of Github. Finally, we have used that model to identify factors that affect language evolution.
\newline
\textbf{Results:} The major findings of the study are: (i) migration, data and data structure are generally the difficult topics of new languages, (ii) the time when adequate resources are expected to be available vary from language to language, (iii) the unanswered question ratio increases regardless of the age of the language, and (iv) there is a relationship between developers' activity pattern and the growth of a language.
\newline
\textbf{Conclusion:} We believe that the outcome of our study is likely to help the owner/sponsor of these languages to design better features and documentation. It will also help the software developers or students to prepare themselves to work on these languages in an informed way.

\end{abstract}

%% file: Sections/Introduction.tex
\section{Introduction}
\label{sec:introduction}
New programming languages are being introduced to make software development easy, maintainable, robust, and performance-guaranteed~\cite{Maloney2010,pierce2002}. For example, Swift was introduced in June 2014 as an alternative to Objective-C to achieve better performance. At the initial stage of its lifetime, a programming language is likely to have constraints of resources, and consequently, developers using these languages face additional challenges~\cite{Kushida2015}. Naturally, the developers seek help from community experts of question-answering (QA) sites such as Stack Overflow (SO). Hence, it is expected that the discussions on issues related to a new language in SO represent the different characteristics of the growth of that language and also reflect the demands of the development community who use that language.

After the release of a new programming language, it takes time for the developers to get acquainted with that language. Earlier releases of new languages often contain bugs. The developers who work on the new languages are likely to face problems that are similar to the solved problems of mature languages. Developers of the new languages often feel the absence of a library or feature that has already been available in other languages. Therefore, the discussions on a new language are likely to differ from that of a mature language. To the best of our knowledge, there is yet to be any software engineering research that focuses on the specific characteristics of the new languages by mining relevant discussions from SO.

In this study, we would like to fill this gap, analyzing the discussions on Swift, Go, and Rust that are the most popular programming languages introduced after the inception of SO (2008). Our study is limited to these three languages because other new languages have very small footprints in SO. Being born after SO, the evolution of these languages, right from the beginning, is expected to be reflected in SO. From now on, by the \emph{new language}, we imply Swift, Go, and Rust languages. We also match the SO discussions with the relevant activities in GitHub in the required cases.

The primary goal of this research is to study how software developers
discuss and support three new programming languages (Go, Swift, Rust) in Stack
Overflow. On this goal, we conduct two studies:
(1) Understanding New Language Discussions: We aim to understand what topics developers 
discuss about the three new programming languages, whether and how the topics are similar and/or different across 
the languages, and how the topics evolve over time. (2) Understanding New Language Support: We aim to understand  
what difficulties developers face while using the three new languages, and when and how adequate resources and expertise 
become prevalent to support the three new programming languages in Stack Overflow. In particular, we answer five 
research questions around the two studies as follows. 

\begin{itemize}[leftmargin=10pt]
  \item \textbf{Study  1. New Language Discussions}: We answer two research questions:
  \begin{enumerate}[label={\textbf{RQ\arabic{*}.}}, leftmargin=30pt]
    \item \textbf{What topics are discussed related to Swift, Go, and Rust?}
    
    This investigates the discussion topics of the developers of new languages. Identification of the discussion topics may help the sponsor to design a feature roadmap that actually facilitates the requirement of developers.
    
    \item \textbf{How do the discussed topics evolve over time?}
    
    The community’s discussion topics is likely to vary over time, as resources evolve continuously.  This analysis would enable us to investigate any possible relation between discussion topics and real-world dynamics, such as new releases. We found that a new release does not initiate any significant change in the evolution of discussion topics.
  
    \end{enumerate}
  \item \textbf{Study 2. New Language Support}: We answer three research questions:
  \begin{enumerate}[label={\textbf{RQ\arabic{*}.}},start=3, leftmargin=30pt]
    \item \textbf{How does the difficulty of topics vary across the languages?}
    
    Developers of new languages face problems that are rarely answered or get \emph{delayed answers}. By the \emph{delayed answer}, we imply that answer which is accepted by the user and received after the median answer interval of that month. We want to know about these questions so that special measures can be taken to answer this question. We found that questions related to migration, data and data structure are considered as difficult topics in all three languages.
    
    \item \textbf{When were adequate resources available for the new programming languages in Stack Overflow?}
    
      In this research question, we want to know the time interval, after which we can expect the availability of these resources of new languages in Stack Overflow at a satisfactory level. The use of programming languages is significantly related to the availability of resources of those languages. This question will help developers to make design decisions related to software development. We have seen that two years after the release, sufficient resources can be expected for Swift, whereas this period is three years for Go. We have also found the evidence of having an inadequate resource of Rust language in Stack Overflow.

    \item \textbf{Is there any relationship between the growth of the three programming languages and developers' activity patterns?}
    This question investigates the relationship between developers' activity  (e.g.,  question,  answer)  and the growth of a  language. Language projects maintain a Github repository that supports feature requests~\citep{Bissyande2013} and bug reports through  Github issues.  We used those issues as an indirect measure for language growth. We found evidence of  relationships between developers' activity and the growth of a language.
    
    \end{enumerate}

\end{itemize}

Our findings show that questions related to ``migration" are common among new languages. To facilitate developer efforts, platform owners should provide detailed documentation of steps to migrate from conventional sources. In this study, we identified the duration, after which adequate resources become available in SO. This finding can help developers to make any decisions regarding migration to a new language. In addition, language owners should provide support until adequate resources become available in the QA community. Moreover, our study identifies some of the factors that influence the evolution of new languages. The finding can help language owners to prioritize their goals.



A preliminary version of this paper appeared previously as a short conference
paper~\cite{Chakraborty2019}. The only overlap between the previous
paper~\cite{Chakraborty2019} and the current paper is Research Question 4, i.e.,
`When were adequate resources available for the new programming languages in
Stack Overflow?'.

\noindent\textbf{Paper Organization.} The rest of the paper is organized as follows. Section~\ref{sec:Background and Data Collection} describes the background of our study and the data collection procedure. Section~\ref{sec:Dev discussions} reports the research questions about developers' discussion. Section~\ref{sec:Dev support} presents the research questions about the developers’ support to the three new languages. Section~\ref{sec:Implication} discusses the implications of our findings. Section~\ref{sec:validity} discusses the threats to validity. Section~\ref{sec:Related Works} presents the related work to our study, and Section~\ref{sec:conclusion} concludes the paper.

%% file: Sections/Background.tex
\section{Background and Data Collection}
\label{sec:Background and Data Collection}
\subsection{Stack Overflow Q\&A Site}

Q\&A sites have become very popular in recent years. There are several Q\&A sites that programmers use to ask questions, solve problems they encounter, provide answers to other people's problems, and discuss different approaches. Stack Overflow is the most popular of these sites. Since its inception in 2006, it has become a popular and reliable platform for sharing knowledge among programmers. As a result, Stack Overflow has plenty of resources for programmers on a variety of topics. From the beginning to 2020, 49,53,854 developers have asked 2,01,28,125 questions on 59,524 different topics in the stack overflow.

\subsection{New Programming Languages Discussions in Stack Overflow}

About 37~\citep{wiki:Timeline} programming languages have been released after the inception of Stack Overflow in 2008. Most of the new languages (released after 2008) have a little footprint in SO, which is insufficient to formally analyze the interaction between developers and programming languages. For selecting the language, we have used the SO survey~\citep{StackoverflowSurvey} and the newly released language list~\citep{wiki:Timeline}. In Table~\ref{table:language stats}, we show the footprints of the languages in SO. To do a comparative analysis of the evolution with three new languages, we picked one high footprint language (Java) and one medium footprint language (Python). Javascript has the highest footprint, but it is primarily used for web clients only. We have selected Java as the representative of top-tier languages due to its wide range of use. We have selected Python as the representative of mid-tier languages due to its recent emergence.
\input{Tables/language_stats.tex}

\subsection{Data Collection}
\begin{figure}[t]
\centering
\includegraphics[scale=0.7]{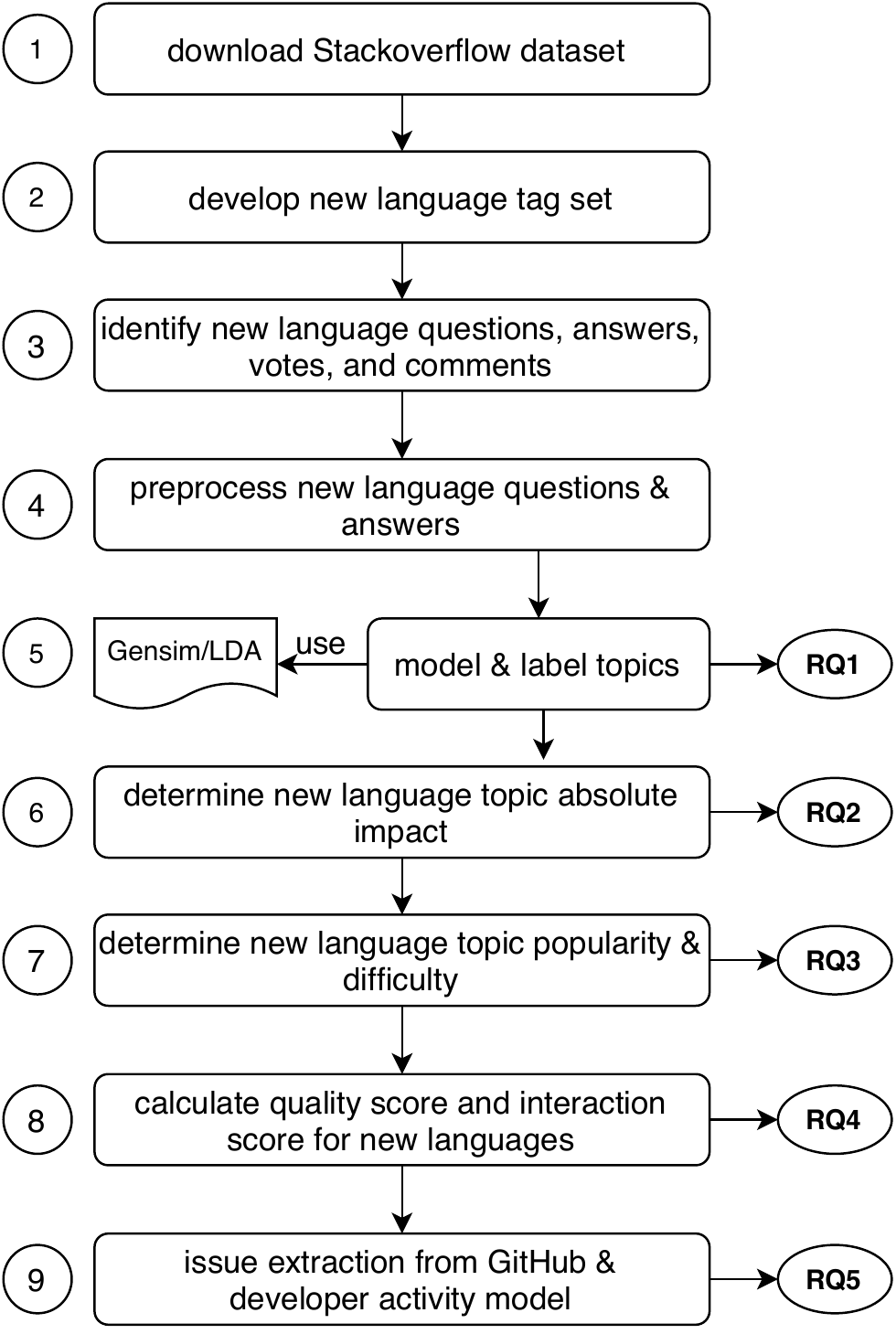}
\caption{An overview of the methodology of our study}
\label{fig:methodology}
\end{figure}

The following steps are carried out to develop the dataset for this study:
\begin{enumerate}[leftmargin=15pt,itemsep=0pt]
    \item We download the SO dataset,
    \item We identify list of tags related to the three languages in SO,
    \item We extract all questions and accepted related to the list of tags from SO, 
    \item We extract issues reported to the GitHub repositories of the three languages.
\end{enumerate}
Figure \ref{fig:methodology} shows an overview of the methodology of our study. We explain the steps below.
\subsubsection{Download Stack Overflow dataset} For our analysis, we have collected the January 2018 Stack Overflow data dump, which is available in the Stack Exchange data dump. In Stack Overflow schema, both question and answer are considered as \emph{post}. The post table of the data dump contains all the information of a post like a title, tags, body, creation date, view count, type (question or answer), and accepted answer identifier. An answer is accepted if the questioner marks that answer as accepted. Our dataset includes 4,17,82,536 questions and answers posted over 9 years from August 2008 to January 2018 by 39,40,962 users of Stack Overflow. Among these posts 1,63,89,567 (39\%) are questions and 2,52,97,926 (61\%) are answers of which 87,04,031 (21\%) are marked as accepted answers.
 
\subsubsection{Develop tag set} To compare the growth of languages, we have to separate the posts by language. Posts on Stack Overflow can be about any topic, and we need a way to identify posts by language. Every Stack Overflow post is associated with at least one tag. We consider a post associated with one of the new languages if that contains at least one tag of the respective language tag. \ra{We have created an initial set of tag $\uptau_0$ for each of the languages. One of the authors checked the initial tagset. Like Vásquez et al. [2], we scaled down the full Stack Overflow (SO) tag set by performing a wildcard query (e.g.; ”SELECT * FROM Tags WHERE TagName like '\%swift\%' order by Count desc”). After that, the search space becomes feasible to perform a manual inspection. The initial tagset is available at \href{https://git.io/JTIqL}{GitHub}.} Next, we go through the Stack Overflow dataset $\mathcal{S}$ and extract questions $\rho$ whose tags contain a tag from $\uptau_0$. Third, we extract tags of the posts in $\uprho$ to form the set of candidate tags $\uptau$. Now we have a set of tags $\uprho$ for each language, which includes all tags of that language. However, set $\uprho$ may include tags that may be irrelevant to new languages. Hence, following the approach of Rosen et al.~\citep{Rosen2015}, we have used two heuristics, $\alpha$ and $\beta$, to find the significantly relevant tags for each language.

    \begin{equation}
        \alpha = \dfrac{number \ of \ posts \ with \ tag \ t \ in \ \uprho}{number \ of \ posts \ with \ tag \ t \ in \ \mathcal{S}}
    \end{equation}
    \begin{equation}
         \beta = \dfrac{number \ of \ posts \ with \ tag \ t \ in \ \uprho}{number \ of \ posts \ in \ \uprho}
    \end{equation}

We have experimented with a broad range of $\alpha$ and $\beta$  and found that
$\alpha = 0.01$ and $\beta=0.01$ provides a significantly relevant set of tags.
The values are consistent with previous research on finding tags for big data or
concurrency tags~\cite{Bagherzadeh2019, Ahmed2018}. The tag set used to extract posts in this study is available at  \href{https://git.io/JTIqL}{GitHub}.

Our new language tag set is extensive and covers a large spectrum of tags related to new languages. The name of language or language versions is a highly relevant tag to identify posts. \emph{Swift 2.1, go,} and \emph{rust} are this kind of tag in our tag set. In terms of relevance, the next type of tag is is the name of a library/framework with or without a version. As these libraries/frameworks are applicable to a particular language they can be used to identify the post of that language. These types of tags in our tag set are \emph{grails2.0, beego, cocoa, rust-tokio}, etc. The third type of tag is named after a specific feature of the language. \emph{Goroutine, unmarshalling, traits} are such kind of tag in our tag set. In addition to highly focused tags, our tag set also includes generic tags such as \emph{concurrency, protocol}.


\subsubsection{Extract posts of new languages} 
Using the tag set prepared in the previous step, we have separated the posts by language. We have 4,37,880 Swift posts, consisting of 1,88,065 (43\%) questions and 2,49,815 (57\%) answers of which 94,310 (21.6\%) are accepted answers. We have 72,843 Go posts, consisting of 30,286 (41.6\%) questions and 42,557 (58.4\%) answers of which 19,178 (26.3\%) are accepted answers. We have 18,311 Rust posts, consisting of 8,083 (44.1\%) questions and 10,228 (55.9\%) answers of which 5,964 (32.6\%) are accepted answers.

\subsubsection{Preprocess new language post set} In this step, posts of the new language are preprocessed to reduce noise.  The preprocessing steps include the removal of code segments, HTML tag, and URL, exclusion of stop words (e.g.,  a, the,  is), and word stemming.   We have used porter stemming for converting the word into their root form.

\subsubsection{Model and label new language topics} In this step, we use the Gensim implementation of latent Dirichlet allocation (LDA)  to identify new languages' topics. Previous studies have pointed that LDA topics may be subject to change if the order of documents is changed. Thus we have used a differential evolution algorithm to select LDA parameters. This approach has made our topics more stable. After extracting the topic, we manually labeled the topics with an appropriate title.

\subsubsection{Calculate topic absolute impact} The absolute topic impact shows the absolute proportion of a particular topic in a particular month’s posts. In this step, the absolute impact of topics are calculated using equation \ref{eq: absolute impact}.

\subsubsection{Calculate topic popularity \& difficulty} In this step, for each topic, we have extracted the number of posts without accepted answers and the median time to answer. Later we have calculated the correlation between without accepted answer percentage and the median time to answer.

\subsubsection{Calculate quality score and interaction score} We have calculated the quality score and interaction score for each language using equations \ref{eq: quality score} and \ref{eq: interaction score}. After that,  we used the scores to determine the stable point date,  a date after which sufficient resources would be available in the SO.
\begin{figure*}[t]
    \centering
    \subfloat[\# of Repositories Checked from Github]{{\includegraphics[scale=0.25]{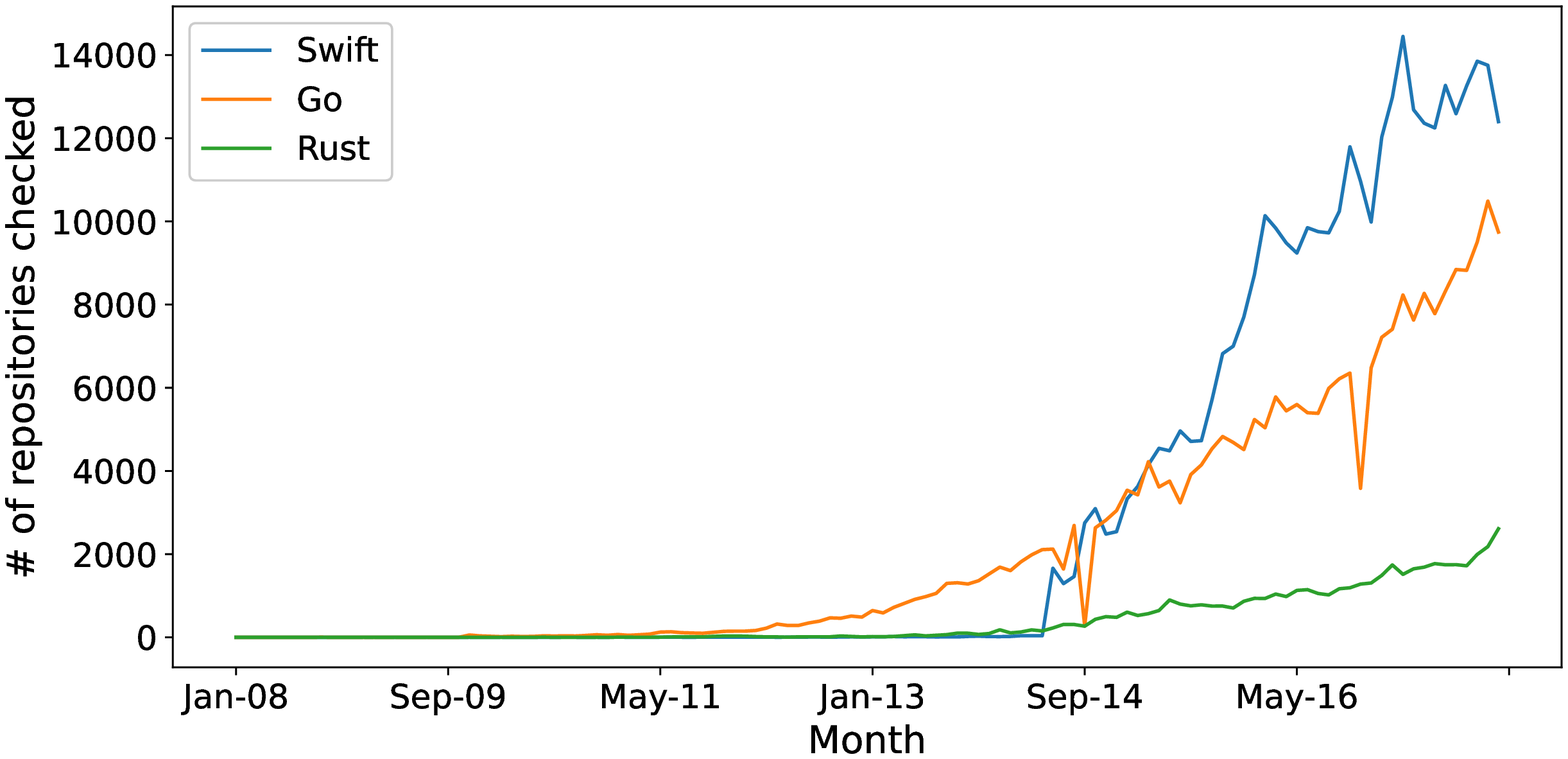} }}%
    \subfloat[\# of Users Checked from Github]{{\includegraphics[scale=0.25]{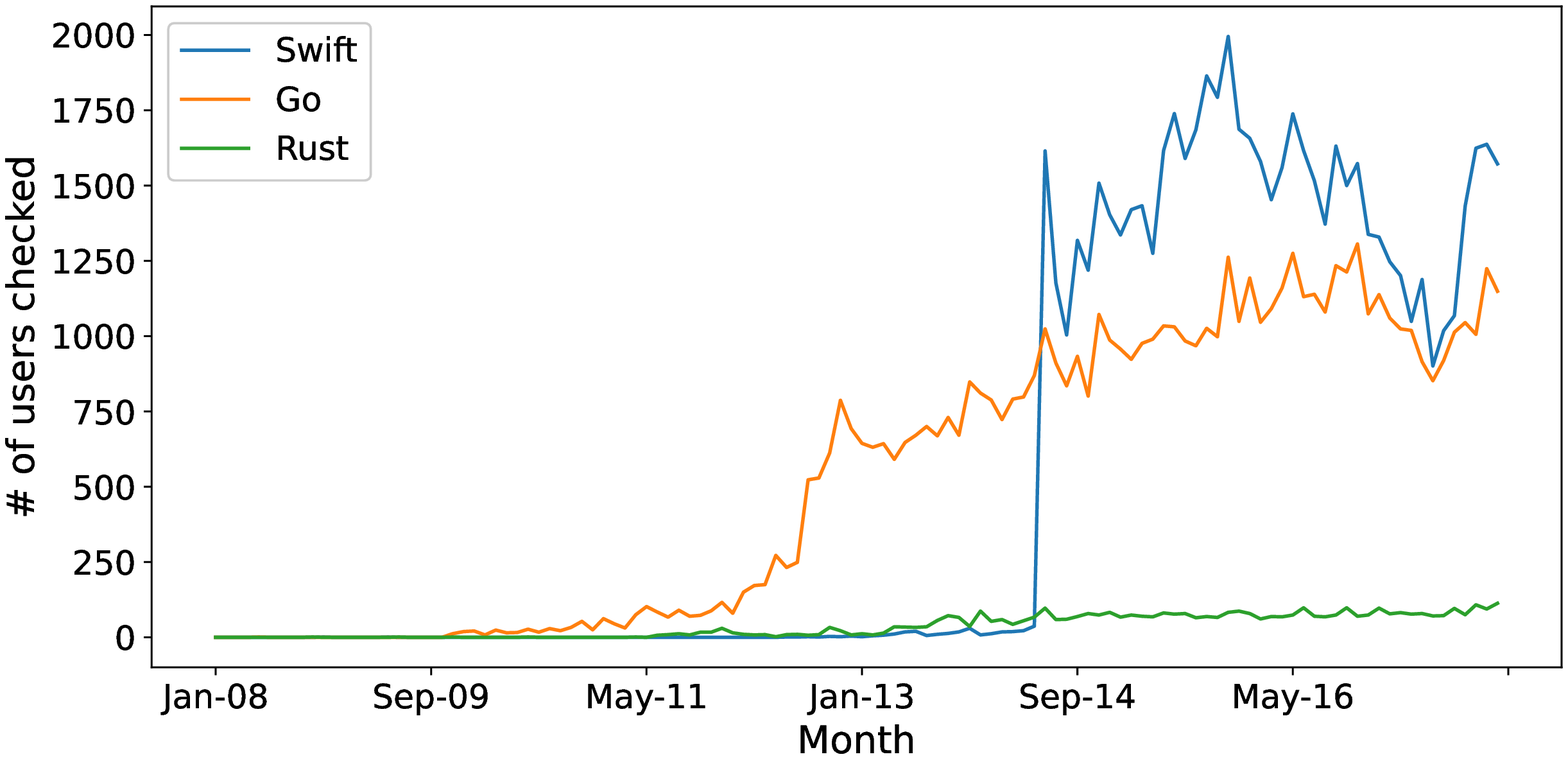}}}%
    \enskip
    
    \caption{\# of Repositories and Users checked from Github}%
    \label{fig:Github Data}
\end{figure*}

\subsubsection{Data extraction from GitHub \& model developer activity} GitHub provides access to data of public repositories and users through its \href{https://developer.github.com/v3/}{public API}. The new languages have their \emph{official repository} in GitHub. In this step, we have collected the creation date and closing date of all the issues from these new languages' official repositories. GitHub issues have two states, `open' and `closed.' As soon as an issue is taken care of, it changes its state from `open' to `closed.' We collected states and frequencies of the issues. \ra{Then for each month, we collected the number of repositories and users of each new language. Github sets the dominant programming language of a repository as the language of that repository. We have used Github Search for collecting repository and user count. Using that specific search language, one can search and count the repositories of a particular language. We have queried for each month and each of the languages. In the search process, we have excluded the fork repositories. Another part of Github data was users. To collect user data, we have collected all the repositories of a particular language, and then we collected the unique committers of those repositories. After finding those committers, we collected their joining date and counted the number of users joined each month. The number of repositories and users we have checked are presented in Figure~\ref{fig:Github Data}. Then we used regression to model developers' activity.}

%% file: Tables/language_stats.tex
\begin{table}
\centering
\caption{Stack Overflow footprint of programming languages.}
\begin{threeparttable}
\begin{tabular}{|l|r|r|r|r|}
\hline
\textbf{Language}   & \textbf{Total Post Count}   \\ \hline
Javascript & 4875127(10.6399\%) \\ \hline
Java       & 4101937(8.9524\%)  \\ \hline
Php        & 3515748(7.673\%)   \\ \hline
C++        & 2575423(5.6208\%)  \\ \hline
Swift\tnote{*}      & 538542(1.1754\%)   \\ \hline
Python     & 276022(0.6024\%)   \\ \hline
Go\tnote{*}         & 91460(0.1996\%)    \\ \hline
Rust\tnote{*}       & 23850(0.0521\%)    \\ \hline
Kotlin\tnote{*}     & 5936(0.013\%)      \\ \hline
Dart\tnote{*}       & 3991(0.0087\%)     \\ \hline
Ballerina\tnote{*}  & 403(0.0009\%)      \\ \hline
\end{tabular}
\begin{tablenotes}
\item [*] released after 2008

\end{tablenotes}

\end{threeparttable}
\label{table:language stats}
\end{table}

%% file: Sections/Dev_Discussions.tex
\section{Developers' Discussions about the Three New Languages}
\label{sec:Dev discussions}

\input{Sections/RQ1}

\input{Sections/RQ2}

%% file: Sections/RQ1.tex
\subsection{RQ1. What are the topics of discussions related to Swift, Go, and Rust?}
\label{RQ1}
\subsubsection{Motivation} 
In this work, we explore the SO footprint of three new languages introduced after SO and become popular in the developer community for knowledge sharing. Hence, it is likely that the issues developers face while working with these languages will be reflected in the posts and discussions on these languages in SO. If the queries are organized according to topics and characteristics of the responses are analyzed accordingly, it would be helpful for the sponsors of those languages. The lack of resources (such as proper documentation) will be revealed for the most visited topics, and the relevant people may address those in an organized way. Hence, our first research question is intended to analyze these languages' discussions by dividing these into different topics and categories.

\subsubsection{Approach.} We used LDA to identify the topics of developers' discussion. LDA (Latent Dirichlet Allocation) is a generative statistical model commonly used for topic modeling~\cite{Blei2003}. In LDA, it is assumed that each document is a mixture of a small number of topics. It is believed that LDA topics are related to the order of documents in the dataset. If the order is changed, the topics will likely be changed. We have calculated the Raw score for LDA topics to mitigate the risk. The raw score is a modified version of the Jaccard Index. In calculating the raw score, the LDA parameters are kept constant, and the data orders are changed. This process is repeated 10 times to identify the raw score of LDA for one set of parameters. To ensure topic stability and determine the best LDA Parameters, we filled up the LDA parameters with a set of arbitrary values and calculated the raw score for each group. Using the best (in terms of raw score) parameters of the current run, the second run starts. This process has been continued for three generations, and we received the best stability ensuring the LDA parameter. Next, using those parameters, we run the LDA. From LDA, we received a set of keywords for each topic. To label the topics, we identified the dominant topic of all the posts. Next, for each topic, we continued the following labeling strategy. First, we randomly selected twenty posts, where the dominant topic is this topic. Then we manually analyzed the topic keywords along with the posts and labeled the topics. The first author merged the topics into higher level categories first. Then it was reviewed by the second and third authors. The issues identified were resolved by a detailed discussion involving all the authors. Furthermore, we have extracted some of the features of the adopted topics in prior works~\cite{Ahmed2018, Nadi2016, Bagherzadeh2019} to measure popularity.
\begin{itemize}[leftmargin=10pt]
\item\textbf{Average view of post}: SO collects view counts for each post. Using this metric, we can get an indication of the public interest. The intuition is that if many developers view a post, then this post is either very popular or the problem is common among the developers of new languages. For this reason, we have collected average views for each topic.

\item \textbf{Average number of post count as favourite} In SO, users can mark a post as a personal favorite if the post is helpful. Favourite facility notes things that are important or interesting to developers. Developers can return to their favourite posts from the favourite tab in Stack Overflow. We collected the average favorite count for each topic in the new language. The metric will reveal how helpful/aligned the posts are with the developers' goals.

\item \textbf{Average Scores}: In SO, an interesting/unique question or best solution can be rewarded by upvote. Where the attribute “favourite” expresses developers' individual choice, upvote tells the fellow developers whether the post is useful or not. SO then aggregates the votes (summation of the upvotes less than the summation of the downvotes) and presents them as scores. In this study, we summed up all posts' scores and divided them by the number of posts to calculate each topic's average score. This score  of each topic will be used as metrics of perceived collective values.

\end{itemize}

\input{Tables/raw_score}
\input{Tables/coherence_score}
\input{Tables/topic_category_popularity_swift}
\begin{figure}[t]
\centering
\includegraphics[scale=0.25]{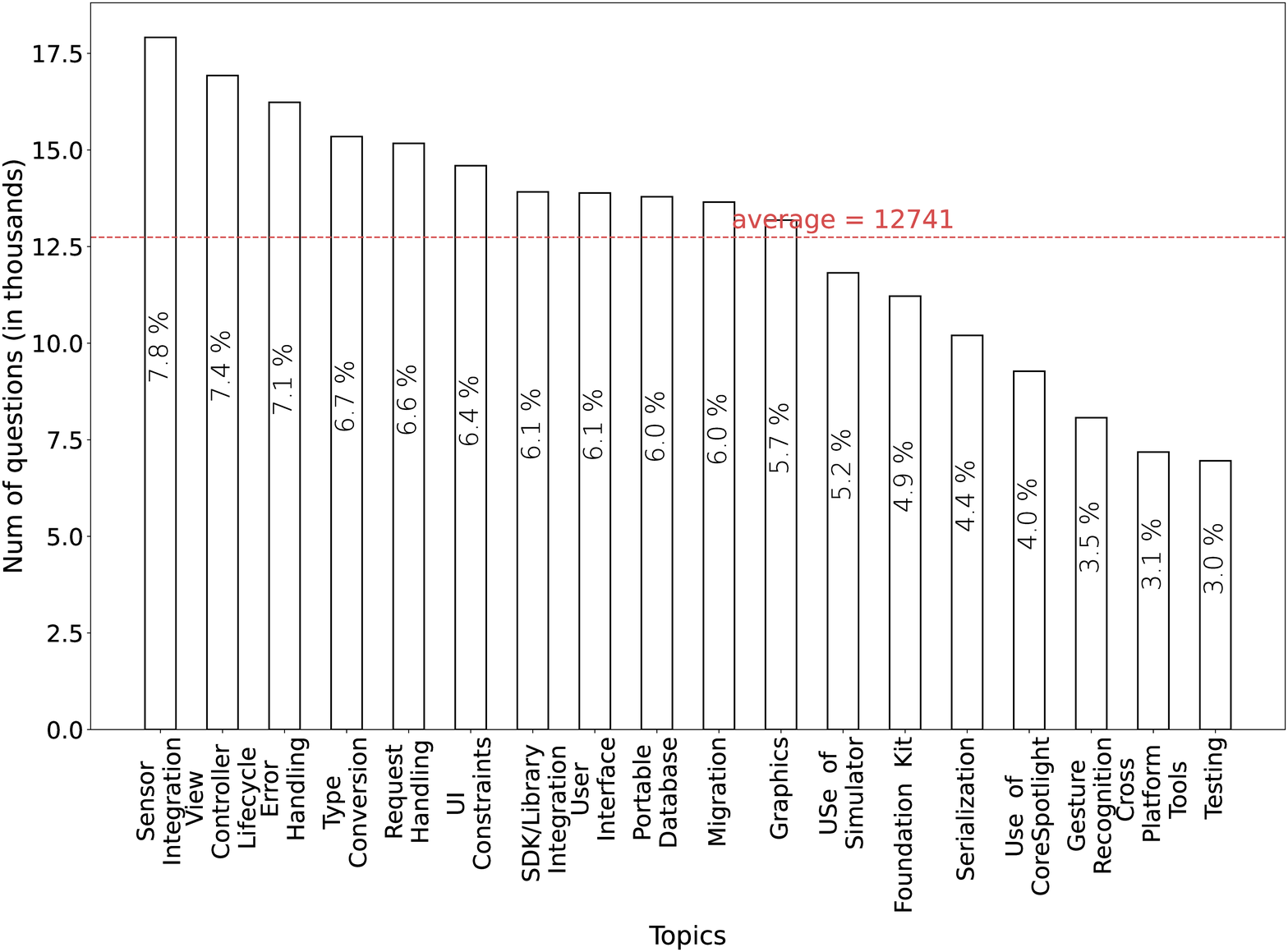}
\caption{Swift topics and number of their questions.}
\label{fig:Swift topic question count}
\end{figure}

\input{Tables/topic_category_popularity_go}
\begin{figure}[t]
\centering
\includegraphics[scale=0.25]{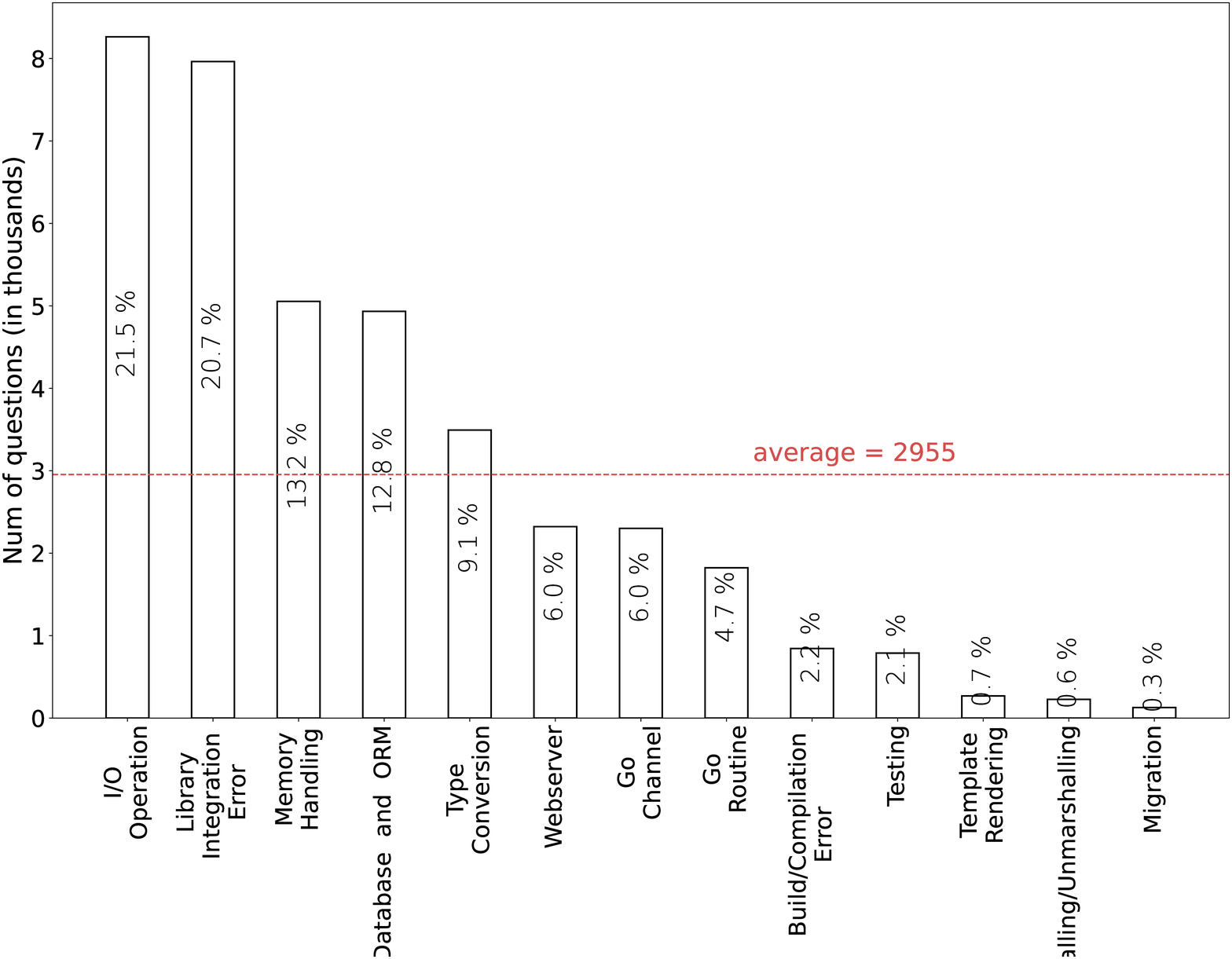}
\caption{Go topics and number of their questions.}
\label{fig:Go topic question count}
\end{figure}
\input{Tables/topic_category_popularity_rust}
\begin{figure}[t]
\centering
\includegraphics[scale=0.25]{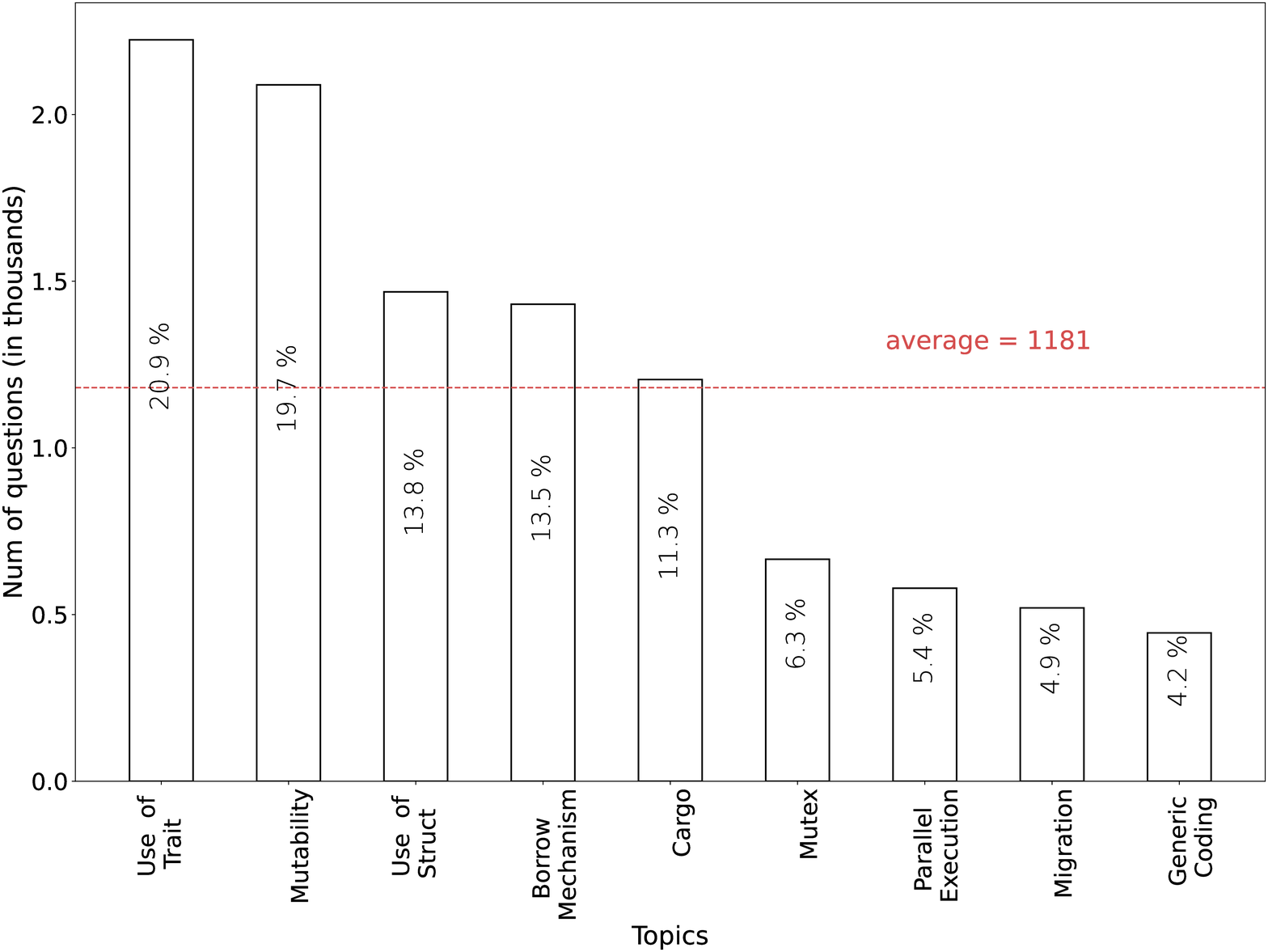}
\caption{Rust topics and number of their questions.}
\label{fig:Rust topic question count}
\end{figure}

\subsubsection{Result.} The raw score of our approach presents the stability of the topics identified by LDA. The higher the raw score, the higher the stability.  In this study, we have used three words while calculating the raw score.  The raw score achieved in our study is presented in Table~\ref{table:raw score}. We also calculated the topic coherence score for our LDA model.  The topic coherence score measures the quality of the extracted topics~\cite{Syed2017}.   The higher the coherence score,  the higher the quality of topics. The topic coherence score is presented in Table~\ref{table:cohernece score}.  Table \ref{table:topic_popularity_swift}, \ref{table:topic_popularity_go}, and \ref{table:topic_popularity_rust} shows the discussion topics for each language sorted in terms of average views. It also shows the number of posts related to each topic and its popularity through the popularity metrics: views, likes, and scores received from users in Stack Overflow.


\subsubsection*{\textbf{Swift Topics}}
The percentage of Swift posts related to each topic is presented in Figure ~\ref{fig:Swift topic question count}. From Table ~\ref{table:topic_popularity_swift} we can observe that 5 of the 18 topics of Swift are related to UI. They are User Interface, View Controller Lifecycle, UI constraint, Gesture Recognition and Graphics. These topics include questions like, a) how a specific UI functionality can be achieved, b) how to use multiple UI components together, c) how the life cycle of view controller that manages applications UI interface changes, d) using 2D and 3D graphics components for game development, etc. More than 17\% of the Swift related posts are on these topics. An example of posts under this topic is a developer asking on Stack Overflow, ``I am making a game on Xcode using sprite kit it is, and I need to add an angry bird slingshot like to the ball, I don’t know how can I apply it." Swift is mainly used to develop iOS applications and games with state-of-the-art user interfaces. As a result, UI related posts are generally higher for Swift. 

3 of the 18 topics of Swift are related to Data and Data Structure. They are Data Handling, Type Conversion, Mutability, and Database. These topics include questions like, a) how to save, stream or receive video data from the network, b) how to perform custom data type conversion and how to write proper syntax for typecasting, c) the use and syntax of immutable data, d) how to perform CRUD and other data manipulation operation in the portable database and corresponding framework provided by Swift, etc. More than 17\% of the Swift related posts are on these topics, and more than 6\% of them are related to data type conversion. The posts related to data type conversion have the highest average score (3.1) among the Swift posts, meaning the posts' answers are generally helpful to the developers. An example of posts under this topic is a developer asking on Stack Overflow, ``I am bit confused with typecasting in Swift. Have a small doubt. What is the difference between `as?', `as!' and only `as.' And can we say `as' is similar to `is'".

About 6\% of the Swift related posts are related to Migration. This topic includes questions for two types of migration problems, a) when the developers face problems to recreate something in Swift while migrating from another language, mostly from Objective-C, and b) when the developers face problems in XCode while migrating from an object project. An example of posts under this topic is a developer asking on Stack Overflow, ``I’m not sure if this is something to do with SWIFT or some bug, but I used to be able to call this in objective c". Objective-C is the predecessor language of Swift, and migration is quite common between these two. Xcode is an IDE for Swift that generally introduces a new major version every year, and migration from an older version to the newer is quite common as well.

About 6\% of the Swift related posts are on the remaining topic, Library/SDK. This topic includes questions related to Swift libraries or SDK primarily on the Foundation Kit. The Foundation Kit or just Foundation is an Objective-C framework that provides basic classes such as wrapper classes and data structure classes with a fixed prefix NS. It is part of the Swift standard library. An example of posts under this topic is a developer asking on Stack Overflow, ``My NSLog result shows that the string is there. However I get the error mentioned in the title of this question. When I replace the string `resortName' with `location' and store the whole object instead the error goes away." The Foundation Kit is a fundamental framework that is quite old and mature, so the number of questions (posts) will likely be lower.
\boxtext{Swift users mostly discussed about application.}

\subsubsection*{\textbf{Go Topics}}
The percentage of Go posts related to each topic is presented in Figure~\ref{fig:Go topic question count}. From Table~\ref{table:topic_popularity_go} we can observe that 3 of the 13 topics of Go are related to Data and Data Structure. They are Database, Type Conversion, Marshalling/Unmarshalling, and Go Channel. These topics include questions related to, a) problems in using pointers, slicing, errors related to reference or pointer, b) custom types, type conversion, typecasting in Go, c) convert Go object/struct to JSON (marshalling), convert JSON to struct (unmarshalling), pointer marshalling/unmarshalling, d) proper usage of Go channel through which goroutines communicate strictly type data, etc. More than 22\% of the Go related posts are on these topics, and more than 16\% of them are related to Unmarshalling/Marshalling. The posts related to Unmarshalling/Marshalling have a much higher average score (8.11) among the Go posts, meaning the posts' answers are generally helpful to the developers. An example of posts under this topic is a developer asking on Stack Overflow, ``what the best way is to perform idiomatic type conversions in Go. Basically my problem lays within automatic type conversions between uint8, uint64, and float64. From my experience with other languages a multiplication of a uint8 with a uint64 will yield a uint64 value, but not so in go."

More than 13\% of the Go related posts are related to Memory. This topic includes questions related to problems in memory allocation and sharing. Go supports automatic memory management, such as automatic memory allocation and automatic garbage collection. It is interesting that developers still face issues related to memory. An example of posts under this topic is a developer asking on Stack Overflow, ``I want to make an array of size N in go, but I don’t know what N will be at compile time, how would I allocate memory for it?".

More than 2\% of the Go related posts are related to Build/Compilation. This topic includes questions related to build/compilation problem. Go requires a directory structure for compilation, and it seems that the structure is not clear to developers. An example of posts under this topic is a developer asking on Stack Overflow, ``I noticed the go/ast, go/token, go/parser, etc. packages in the src/pkg/go folder. However, the GCC compiler was based on C files located in src/cmd/gc. My question regards the new go command in Go that builds and runs programs: does this tool depend on the packages I referenced above?". More than 0.3\% of the Go related posts are related to Migration. The posts related to Migration have the second-highest average score (7.4) among the Go posts, meaning the posts' answers are generally helpful to the developers. This topic includes questions related to problem developers face while migrating their solution in a different language to Go. An example of posts under this topic is a developer asking on Stack Overflow, ``We want to rewrite kodingen.com backend with Go, which currently is Java, running as a daemon using Jsvc. I have never touched any C in my life; simple requirements give me hope that I can start using this wonderful language. What would you advise? Is C still better?".

More than 21\% of the Go related posts are related to I/O. This topic includes questions related to all types of I/O operations in Go. An example of posts under this topic is a developer asking on Stack Overflow, ``I’m trying to write a golang program to control mpv via issuing commands to a unix socket. This should cause mpv to quit but nothing happens". More than 20\% of the Go related posts are related to Library/SDK. This topic includes questions related to different libraries, the majority of them on the ORM library named GORM. An example of posts under this topic is a developer asking on Stack Overflow, ``I'm using Go with the GORM ORM. I have the following structs. The relation is simple. One Town has multiple Places and one Place belongs to one Town. How can i do such query?". More than 6\% of the Go related posts are related HTTP. This topic includes questions related to serving HTTP requests in Go. An example of posts under this topic is a developer asking on Stack Overflow, ``What website has some good, up to date resources on using Go HTML/templates, especially in regard to parsing HTML files".
\boxtext{Data and data structure related posts are mostly discussed among Go developers.  They represent 31.2\% of posts of Go language.}
\subsubsection*{\textbf{Rust Topics}}
The percentage of Rust posts related to each topic is presented in Figure~\ref{fig:Rust topic question count}. From Table~\ref{table:topic_popularity_rust} 5of the 9 topics of Rust are related to Data and Data Structure. They are Borrow Mechanism, Use of Trait, Mutability, Use of Struct, and Generic Coding. These topics include questions related to a) use of Rust borrowing mechanism to access data without taking ownership, b) use of Rust trait (similar to the interface in Java) especially by new developers, c) use of immutable variable, d) problems to get the exact behavior from Rust struct and in destructuring a struct, e) use of generic programming that deals with generic data types and use of traits in generic algorithms, f) use of persistent data storage, data iterator, database driver, storing custom object into database, etc. More than 72\% of the Rust related posts are on these topics, and more than 27\% of them are related to Use of Trait. The posts related to the Generic coding have the second highest average score (4.1) among the Rust posts, meaning the posts' answers are generally helpful to the developers. An example of posts under this topic is a developer asking on Stack Overflow, ``I’m trying to learn Rust, I'm wondering if it is possible to declare the reader variable earlier with a generic type ..."

More than 11\% of the Rust related posts are related to Library/SDK, primarily on the Rust package manager, Cargo. This topic includes questions related to package’s dependencies, compilation of the packages, and distribution of packages. An example of posts under this topic is a developer asking on Stack Overflow, ``I am developing a cargo package which has both a library and an executable of the same name in the same directory. How can I specify different dependencies for them both?"

2 of the 9 topics of Rust are related to Parallelization. They are Parallel Execution and Mutex. These topics include questions related to a) parallel execution in Rust, b) use of mutex and lock in a multiprocessing environment. More than 11\% of the Rust related posts are on these topics. An example of posts under this topic is a developer asking on Stack Overflow, ``I am having trouble understanding how to modify Option inside a Mutex. Any idea which Rust concept causes this?"

More than 4\% of the Rust related posts are related to Migration. This topic includes questions related to problem developers face mimicking logic in Rust during migration. An example of posts under this topic is a developer asking on Stack Overflow, ``I am hoping to re-write some parts of a Python project in Rust to speed up things. I am capable of returning complex arrays/structures in Python. And this does not work properly in Rust."

\boxtext{More than half (72.05\%) of the discussed topic among Rust developers is related to data and data structure.}

%% file: Tables/raw_score.tex
\begin{table}[]
\caption{Raw score achieved by our approach.}
\centering
\begin{tabular}{cc}
\hline
\textbf{Language} & \textbf{Raw Score} \\ \hline
Swift    & 0.88      \\ \hline
Go       & 0.88      \\ \hline
Rust     & 0.66      \\ \hline
\end{tabular}
\label{table:raw score}
\end{table}

%% file: Tables/coherence_score.tex
\begin{table}[]
\caption{Coherence score of the topics.}
\centering
\begin{tabular}{cc}
\hline
\textbf{Language} & \textbf{Coherence Score} \\ \hline
Swift    & 0.59      \\ \hline
Go       & 0.56      \\ \hline
Rust     & 0.49      \\ \hline
\end{tabular}
\label{table:cohernece score}
\end{table}

%% file: Tables/topic_category_popularity_swift.tex
\newcolumntype{s}{>{\hsize=.13\hsize}X}
\newcolumntype{d}{>{\hsize=.12\hsize}X}
\newcolumntype{m}{>{\hsize=.14\hsize}X}
\newcolumntype{v}{>{\hsize=.29\hsize}X}
\begin{table}[htb!]
\centering
\caption{Swift topics, categories, and their popularity.}
\begin{tabularx}{\textwidth}{mvsssd}
\hline
\textbf{Category}                                & \textbf{Topic}                     & \textbf{Avg. Views} & \textbf{Avg. Favourite} & \textbf{Avg. Score} & \textbf{\#Posts} \\ \hline
\multirow{6}{*}{Application}            & Request Handling          & 939.87         & 0.41               & 1.17            & 15171 \\
                                        & Use of Simulator          & 985.51         & 0.52               & 1.5             & 11819 \\
                                        & Error Handling            & 1416.35        & 0.41               & 1.56            & 16233 \\
                                        & Testing                   & 1482.49        & 0.65               & 2.33            & 6956  \\
                                        & Cross Platform Tools      & 1486.1         & 0.6                & 1.7             & 7182  \\
                                        & Sensor Integration        & 1534.83        & 0.62               & 1.65            & 17914 \\ \hline
\multirow{3}{=}{Data\& Data Structure} & Portable Database         & 1101.06        & 0.4                & 1.26            & 13790 \\
                                        & Serialization             & 1291.47        & 0.49               & 1.36            & 10201 \\
                                        & Type Conversion           & 2334.98        & 0.81               & 3.33            & 15347 \\ \hline
\multirow{3}{=}{Library/ SDK}            & Foundation Kit            & 1103.47        & 0.45               & 1.24            & 11218 \\
                                        & Use of CoreSpotlight      & 1405.68        & 0.63               & 2.4             & 9274  \\
                                        & SDK/Library Integration   & 1617.3         & 0.75               & 2.93            & 13916 \\ \hline
Migration                               & Migration                 & 1408.09        & 0.73               & 2.69            & 13651 \\ \hline
\multirow{5}{*}{UI}                     & Graphics                  & 695.4          & 0.46               & 1.2             & 13185 \\
                                        & User Interface            & 888.84         & 0.31               & 0.85            & 13887 \\
                                        & View Controller Lifecycle & 1089.92        & 0.37               & 1.08            & 16927 \\
                                        & UI Constraints            & 1376.63        & 0.42               & 1.36            & 14594 \\
                                        & Gesture Recognition       & 1590.26        & 0.54               & 1.73            & 8071  \\ \hline
\end{tabularx}

\label{table:topic_popularity_swift}

\end{table}

%% file: Tables/topic_category_popularity_go.tex
\newcolumntype{s}{>{\hsize=.13\hsize}X}
\newcolumntype{d}{>{\hsize=.12\hsize}X}
\newcolumntype{m}{>{\hsize=.15\hsize}X}
\newcolumntype{v}{>{\hsize=.29\hsize}X}
\begin{table}[htb!]
\centering
\caption{Go topics, categories, and their popularity.}
\begin{tabularx}{\textwidth}{mvsssd}
\hline
\textbf{Category}                                & \textbf{Topic}                     & \textbf{Avg. Views} & \textbf{Avg. Favourite} & \textbf{Avg. Score} & \textbf{\#Posts} \\ \hline
\multirow{3}{*}{Application}            & Webserver                 & 1245.48        & 0.59               & 1.81            & 2322  \\
                                        & Template Rendering        & 1706.95        & 0.83               & 2.68            & 269   \\
                                        & Testing                   & 2342.8         & 1.17               & 4.43            & 789   \\ \hline
Build Compilation                       & Build/Compilation Error   & 1880.5         & 0.92               & 2.85            & 844   \\ \hline
\multirow{3}{=}{Data \& Data Structure} & Database and ORM          & 1051.49        & 0.35               & 1.36            & 4933  \\
                                        & Type Conversion           & 2223.41        & 0.91               & 3.3             & 3493  \\
                                        & Unmarshalling/ Marshalling & 5888.08        & 1.51               & 8.11            & 227   \\ \hline
I/O                                     & I/O Operation             & 1767.06        & 0.74               & 2.54            & 8263  \\ \hline
Library/ SDK                             & Library Integration Error & 1858.89        & 0.7                & 2.45            & 7963  \\ \hline
Memory                                  & Memory Handling           & 2336.88        & 0.94               & 3.69            & 5054  \\ \hline
Migration                               & Migration                 & 3949.88        & 2.27               & 7.44            & 128   \\ \hline
\multirow{2}{*}{Parallelism}            & Go Channel                & 1222.28        & 0.68               & 2               & 2301  \\
                                        & Go Routine                & 1729.49        & 0.97               & 2.96            & 1823  \\ \hline
\end{tabularx}

\label{table:topic_popularity_go}
\end{table}

%% file: Tables/topic_category_popularity_rust.tex
\newcolumntype{s}{>{\hsize=.15\hsize}X}
\newcolumntype{d}{>{\hsize=.12\hsize}X}
\newcolumntype{m}{>{\hsize=.22\hsize}X}
\newcolumntype{v}{>{\hsize=.25\hsize}X}
\begin{table}[htb!]
\centering
\caption{Rust topics, categories, and their popularity.}
\begin{tabularx}{\textwidth}{mvsssd}
\hline
\textbf{Category}                                & \textbf{Topic}                     & \textbf{Avg. Views} & \textbf{Avg. Favourite} & \textbf{Avg. Score} & \textbf{\#Posts} \\ \hline
\multirow{5}{=}{Data \& Data Structure} & Borrow Mechanism   & 560.88         & 0.32               & 2.81            & 1431  \\
                                        & Use of Trait       & 567.95         & 0.46               & 3.39            & 2224  \\
                                        & Mutability         & 868.59         & 0.4                & 3.36            & 2089  \\
                                        & Generic Coding     & 929.63         & 0.53               & 4.17            & 445   \\
                                        & Use of Struct      & 1050.56        & 0.44               & 3.89            & 1468  \\ \hline
Library/ SDK                             & Cargo              & 782.68         & 0.56               & 3.66            & 1205  \\ \hline
Migration                               & Migration Problem  & 837            & 0.55               & 4.55            & 520   \\ \hline
\multirow{2}{*}{Parallelization}        & Mutex              & 675.14         & 0.36               & 2.81            & 666   \\
                                        & Parallel Execution & 678.57         & 0.41               & 3.2             & 579   \\ \hline
\end{tabularx}
\label{table:topic_popularity_rust}
\end{table}

%% file: Sections/RQ2.tex
\subsection{RQ2. How do the discussed topics evolve over time?}
\subsubsection{Motivation} 
We have the rare opportunity to observe the evolution of the discussion on the issues belonging to different topics for these three new languages from the relevant SO posts. The community's interest is likely to vary on particular topics over time, as resources and surroundings also evolve continuously. Moreover, with evolution, languages introduce/abandon features. These changes might be reflected in the developers' discussions. This analysis would also enable us to investigate how the topics change and any possible relation between topic-wise post frequency and real-world dynamics, such as new releases. 

\subsubsection{Approach} To compare the evolution of discussed topics we used two metrics, \emph{topic popularity} and \emph{topic absolute impact} introduced in prior works~\cite{Wan2019}. These two metrics are applied on the data received from LDA. The definition of the two metrics is presented below.

Let,  ($z_1, z_2,..., z_k$) is the set of topic probability vector and $dominant(d_i)$ is the dominant topic of document. The dominant topic $dominant(d_i)$ is defined as,
\begin{equation}
    dominant(d_i) = z_k:\theta(d_i, z_k) = max(\theta(d_i, z_j)); 1\leq j \leq K
\end{equation}

Now, the \emph{topic popularity} for each topic $z_k$ in the dataset $c_j$ is defined as,
\begin{equation}
    popularity(z_k,c_j) = \frac{|\{d_i\}|}{|c_j|};dominant(d_i) = z_k; 1\leq j \leq K
\end{equation}
and the \emph{absolute topic impact} of a topic  $z_k$ in month $m$ within corpus $c$ is defined as,
\begin{equation}
    impact\textsubscript{absolute}(z_k, m) = \sum_{d_i\in D(m)}\theta(d_i,z_k)
\label{eq: absolute impact}
\end{equation}
where $D(m)$ is the set of posts in month $m$. The \emph{absolute topic impact} shows the absolute proportion of a particular topic in a particular month's posts where the \emph{topic popularity} presents the proportion of a particular topic in the full dataset.

\subsubsection{Results} 
\begin{figure}[htb]
\hspace*{-.8cm}
\includegraphics[scale=0.29]{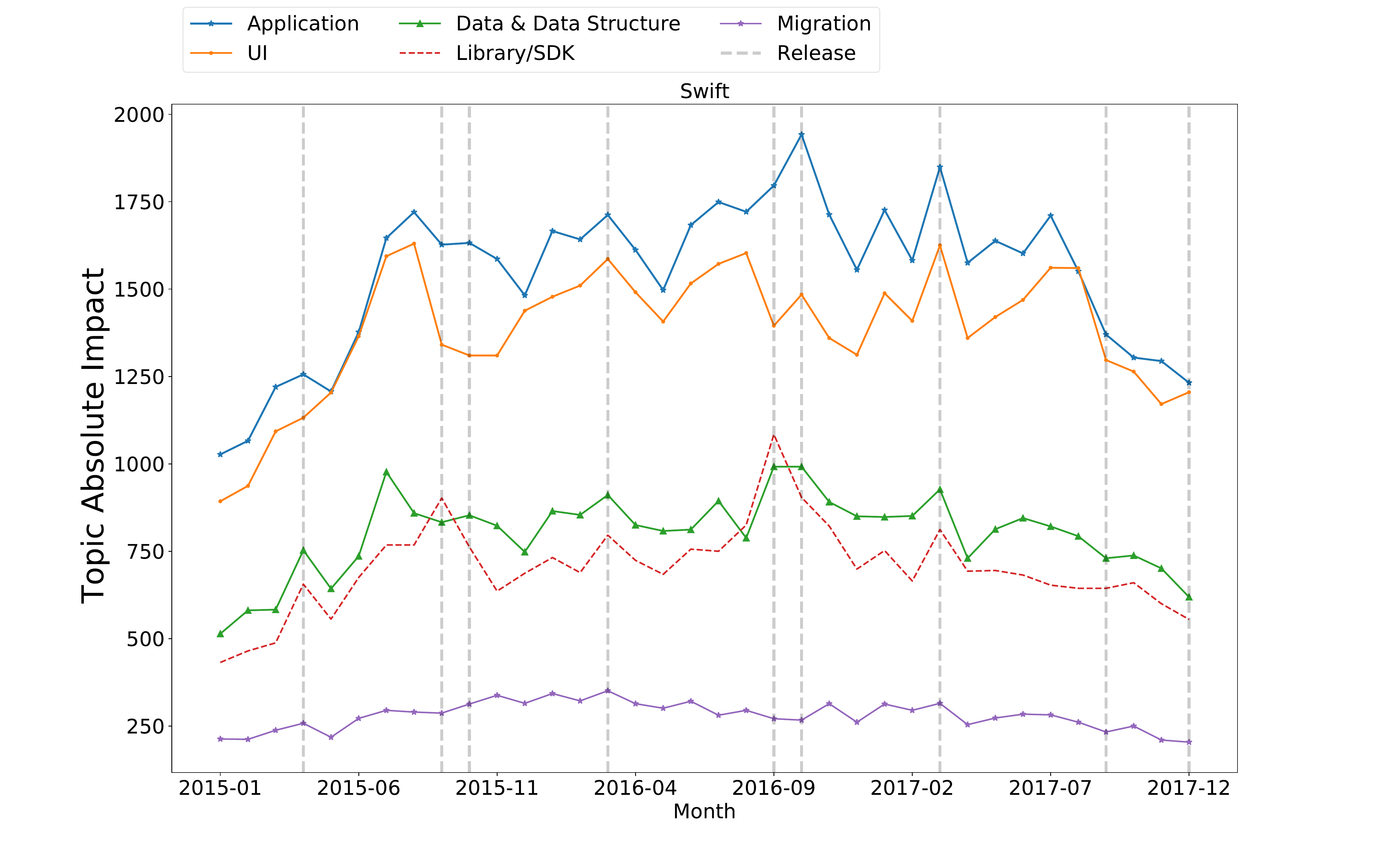}
\caption{Topic absolute impact by the topic categories of  Swift along with release of language version. Each release is a vertical gray dashed line.}
\label{fig:swift topic popularity}
\end{figure}
\begin{figure}[htb]
\hspace*{-.8cm}
\includegraphics[scale=0.29]{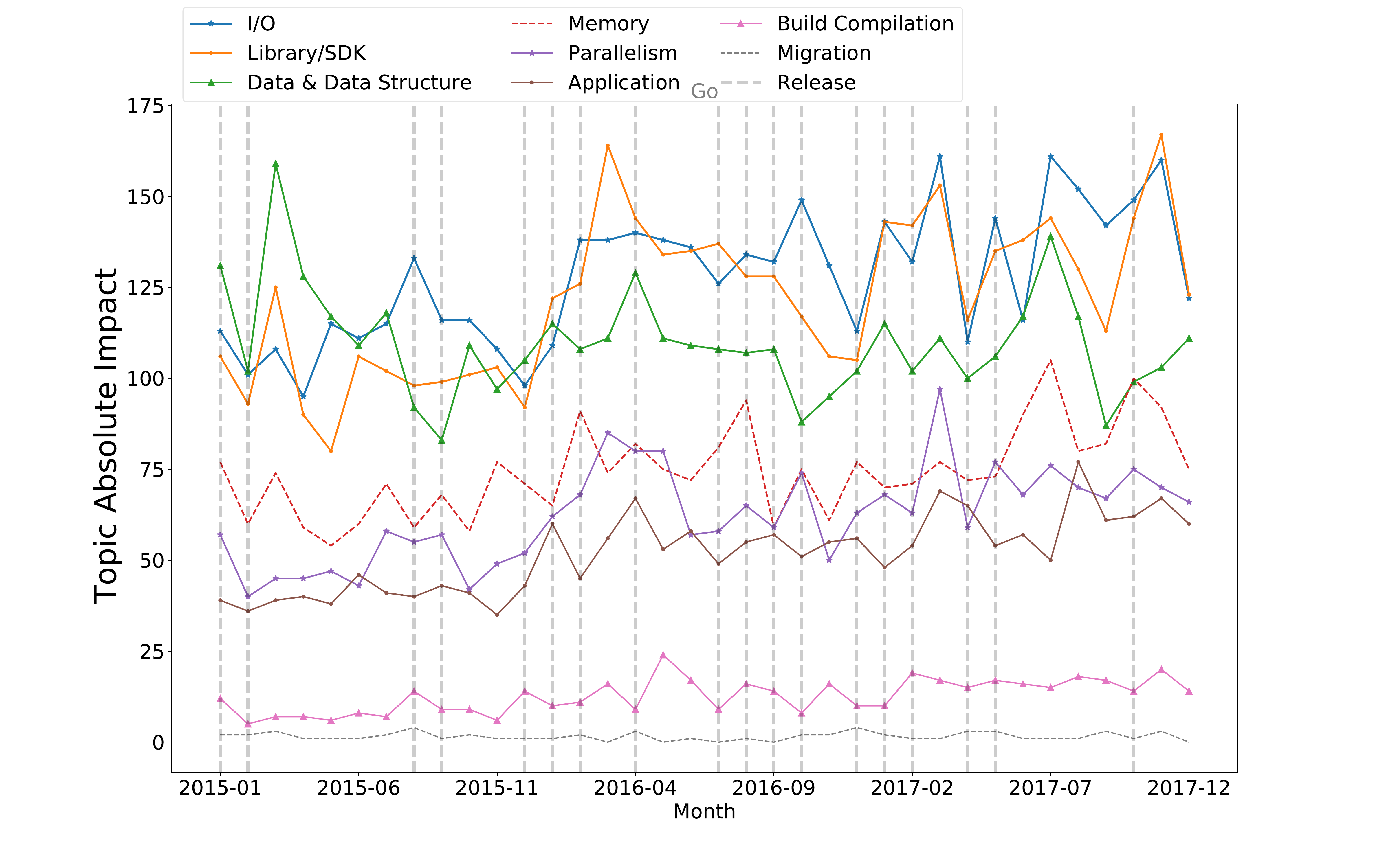}
\caption{Topic absolute impact by the topic categories of  Go along with release of language version. Each release is a vertical gray dashed line.}
\label{fig:go topic popularity}
\end{figure}
\begin{figure}[htb]
\hspace*{-.8cm}
\includegraphics[scale=0.29]{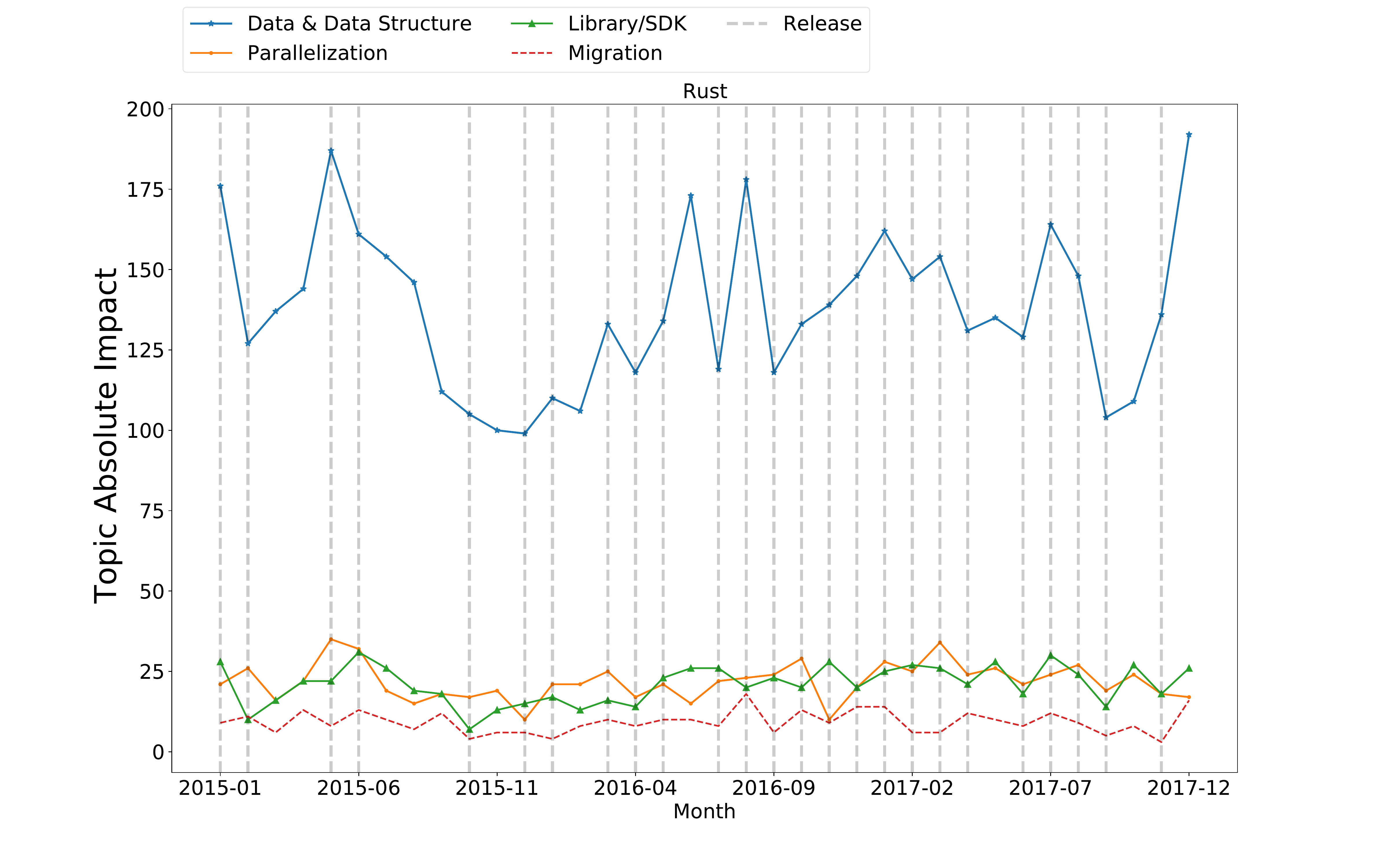}
\caption{Topic absolute impact by the topic categories of  Rust along with release of language version. Each release is a vertical gray dashed line.}
\label{fig:rust topic popularity}
\end{figure}

The topic popularity and absolute topic impact of each topic of each language is presented in Figure \ref{fig:swift topic popularity}, \ref{fig:go topic popularity}, and \ref{fig:rust topic popularity} along with the releases of new version. Both for Go and Rust, the topics related to the category `Data \& Data Structure' remain popular starting from the first day of their discussions in Stack Overflow till the last date of our analysis of the data. For the other language (Swift), however, the topics related to the category `Application' remained the most popular over time, followed by the topics related to `UI'. Most Swift developers are interested in the \emph{user interface} topic. The finding is consistent with real-world observation because Swift is primarily used to create GUI-based software. \ra{Overall, for all topics of Swift, we are observing a downward trend. The release frequency of Swift is comparatively lower than the other two languages. This indicates that Swift developers can have more time than the other languages to learn the specific features offered in a given release. As such, if we normalize the number of questions Swift developers asked per release, it is not surprising that the average number of questions per release for Swift is less than the other two languages. Moreover, in Section ~\ref{RQ4} we have shown that Swift language achieved maturity on November 1, 2016. That means after that point, most of the questions of the Swift are already answered. As most of the questions are already answered in this way, they do not need to ask a new question. As a result, the topic's absolute impact is downward.}


From the topic absolute impact of Go and Rust, the topics related to the `Library/SDK' problem remain the second most discussed throughout the entire timeline. The most commonly discussed library for Rust was Cargo. In Rust `Parallelization' also achieves the second position along with library/SDK topic. There are two main reasons for this. Cargo is Rust's package manager. It is clear from the topic absolute impact of Rust language that Rust developers struggle to use Cargo properly. Rust also does not have a specific guidance on how to do concurrency. This is because Rust simply exposes standard library operating system threads and block system calls like any generic language. The discussions around mutex or parallel execution using Rust in Stack Overflow show the opinions of different developers on the issues and the best practices to handle parallelization in Rust. 



We have anticipated that the release of a major version of the languages may increase the discussion on certain topics. Thus we  have collected the release dates from the official website (for GO) and Github repository (for Swift and Rust) and plotted the topic absolute impact with the release of languages in Figure \ref{fig:swift topic popularity}, \ref{fig:go topic popularity}, and \ref{fig:rust topic popularity}. 
From the figures, we see spikes in the developers' discussions around the release of new version of the three languages at the beginning, i.e., when the three new languages are relatively new to the developers. However, the intensity of such spikes has subsided over time, as the new languages get old. 

\boxtext{The absolute impact is almost constant in all languages except Swift. In Swift, we have noticed a downward trend in the topic absolute impact. On the other hand, the release of a new version of a language does not result in any significant change in the values of topic absolute impact of that particular language as the language gets older.}

%% file: Sections/Dev_Support.tex
\section{Developers' Support to the Three New Languages}
\label{sec:Dev support}

\input{Sections/RQ3}
\input{Sections/RQ4}
\input{Sections/RQ7}

%% file: Sections/RQ3.tex
\subsection{RQ3. How does the difficulty of the topics vary across the languages?}
\label{RQ3}
\subsubsection{Motivation} 
A new language is likely to have some topics with new concepts. Hence, programmers experienced in other languages may find those difficult. Consequently, posts/queries on those topics are likely to have less response from the community. Considering this, we plan to explore those topics for our three languages of interest. The owners/sponsors of these languages can enrich their documentation for these difficult topics with priority. Language instructors will also get an idea of where they should focus more.

\subsubsection{Approach} 

To answer this question we collected two well-known~\cite{Rosen2015,Bagherzadeh2019} metrics for all topics of section \ref{RQ1} to measure difficulty.
\begin{enumerate}[leftmargin=10pt]
\item \textbf{The percentage of posts of a topic without accepted
answers (\% w/o accepted answers)} In SO, if users think that an answer to a query/post provides a solution to that problem, they can mark it as accepted. For each topic, we have collected the average number of accepted answers. Generally, a post is considered difficult if the number of accepted answers is low~\cite{Rosen2015,Bagherzadeh2019}.

\item \textbf{The median time in minutes for an answer to be accepted
(Median Time to Answer (Minutes.)).} We have calculated the median time to get an accepted answer. The more time it takes to get an accepted answer for a post, the more difficult the post is~\cite{Rosen2015,Bagherzadeh2019}.
\end{enumerate}

\subsubsection{Results}
\input{Tables/topic_difficulty_swift}
\input{Tables/topic_difficulty_go}
\input{Tables/topic_difficulty_rust}

Table \ref{table:topic_difficulty_swift}, \ref{table:topic_difficulty_go}, and \ref{table:topic_difficulty_rust}  shows the percentage of questions without an accepted answer and median time (in minutes) to receive an acceptable solution for each of the identified topics in Section \ref{RQ1}. The topics in Table \ref{table:topic_difficulty_swift}, \ref{table:topic_difficulty_go},  and \ref{table:topic_difficulty_rust} are grouped into categories and ordered inside the group based on the percentage of posts without an accepted answer. 

\input{Tables/topic_corr}
To understand the relationship between topic difficulty and popularity, we have performed a correlation analysis. We have chosen the Spearman correlation as it does not assume normality in the distribution of data. Table \ref{table: topic_correlation} shows the correlation. It is clear from Table \ref{table: topic_correlation} that the correlation between the popularity and difficulty metrics is not statistically significant except for the Go language. It seems that difficult topics are not that much popular among Go developers.

\begin{figure}
\centering
\includegraphics[scale=0.5]{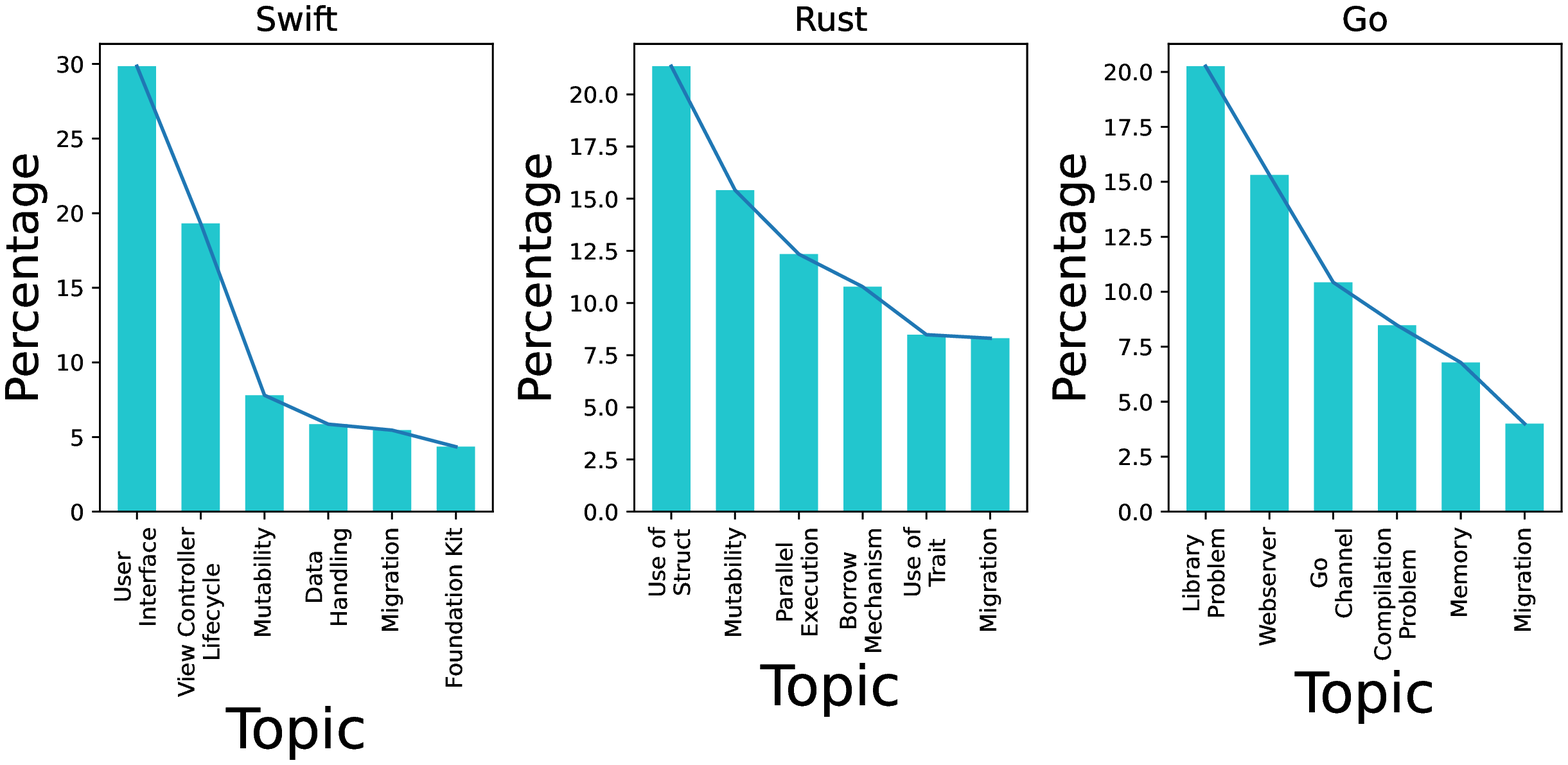}
\caption{Top six difficult topics of new languages}
\label{fig:less_unanswered_topics}
\end{figure}

\boxtext{Difficult topics are not that much popular among Go developers.}

%% file: Tables/topic_difficulty_swift.tex
\newcolumntype{s}{>{\hsize=.15\hsize}X}
\newcolumntype{d}{>{\hsize=.35\hsize}X}
\newcolumntype{m}{>{\hsize=.21\hsize}X}
\begin{table}[htb!]
\centering
\caption{Per topic difficulty of Swift language.}
\begin{tabularx}{\textwidth}{sdmm}
\hline
\textbf{Category} & \textbf{Topic}                     & \textbf{Posts w/o Accepted (\%)} & \textbf{Median Time (m)}          \\ \hline
\multirow{6}{*}{Application}            & Testing                   & 44.11                                 & 30           \\
                                        & Sensor Integration        & 46                                    & 32           \\
                                        & Error Handling            & 46.77                                 & 24           \\
                                        & Cross Platform Tools      & 51.34                                 & 39           \\
                                        & Request Handling          & 59.56                                 & 82           \\
                                        & Use of Simulator          & 62.35                                 & 77           \\ \hline
\multirow{3}{=}{Data \& Data Structure} & Type Conversion           & 35.55                                 & 20           \\
                                        & Portable Database         & 44.11                                 & 32           \\
                                        & Serialization             & 44.74                                 & 31           \\ \hline
\multirow{3}{=}{Library/ SDK}            & Use of CoreSpotlight      & 49.71                                 & 54           \\
                                        & SDK/Library Integration   & 56.93                                 & 111          \\
                                        & Foundation Kit            & 58.6                                  & 63           \\ \hline
Migration                               & Migration                 & 43.77                                 & 35           \\ \hline
\multirow{5}{*}{UI}                     & Gesture Recognition       & 49.06                                 & 38           \\
                                        & UI Constraints            & 49.34                                 & 37           \\
                                        & Graphics                  & 51.83                                 & 104          \\
                                        & View Controller Lifecycle & 52.34                                 & 39           \\
                                        & User Interface            & 52.64                                 & 36           \\ \hline
\end{tabularx}

\label{table:topic_difficulty_swift}
\end{table}

%% file: Tables/topic_difficulty_go.tex
\newcolumntype{s}{>{\hsize=.29\hsize}X}
\newcolumntype{d}{>{\hsize=.23\hsize}X}
\newcolumntype{m}{>{\hsize=.19\hsize}X}
\begin{table}[htb!]
\centering
\caption{Per topic difficulty of Go language.}
\begin{tabularx}{\textwidth}{sdmm}
\hline
\textbf{Category} & \textbf{Topic}                     & \textbf{Posts w/o Accepted (\%)} & \textbf{Median Time (m)}          \\ \hline
\multirow{3}{*}{Application}            & Testing                   & 35.23                                 & 41           \\
                                        & Template Rendering        & 37.17                                 & 45           \\
                                        & Webserver                 & 49.83                                 & 132          \\ \hline
Build/ Compilation                       & Build/ Compilation Error   & 45.14                                 & 69           \\ \hline
\multirow{3}{=}{Data \& Data Structure} & Type Conversion           & 28.4                                  & 24           \\
                                        & Unmarshalling/ Marshalling & 37                                    & 47           \\
                                        & Database and ORM          & 48.63                                 & 129          \\ \hline
I/O                                     & I/O Operation             & 35.68                                 & 50           \\ \hline
Library/ SDK                             & Library Integration Error & 40                                    & 50           \\ \hline
Memory                                  & Memory Handling           & 28.31                                 & 23           \\ \hline
Migration                               & Migration                 & 33.59                                 & 77           \\ \hline
\multirow{2}{*}{Parallelism}            & Go Routine                & 39.11                                 & 48           \\
                                        & Go Channel                & 43.46                                 & 66           \\ \hline
\end{tabularx}
\label{table:topic_difficulty_go}
\end{table}

%% file: Tables/topic_difficulty_rust.tex
\newcolumntype{s}{>{\hsize=.29\hsize}X}
\newcolumntype{d}{>{\hsize=.23\hsize}X}
\newcolumntype{m}{>{\hsize=.14\hsize}X}
\begin{table}[t]
\centering
\caption{Per topic difficulty of Rust language.}
\begin{tabularx}{\textwidth}{sdmm}
\hline
\textbf{Category} & \textbf{Topic}                     & \textbf{Posts w/o Accepted (\%)} & \textbf{Median Time (m)}          \\ \hline
\multirow{5}{=}{Data \& Data Structure} & Generic Coding     & 24.04                                 & 39           \\
                                        & Mutability         & 24.22                                 & 37           \\
                                        & Use of Trait       & 26.75                                 & 44           \\
                                        & Borrow Mechanism   & 26.97                                 & 41           \\
                                        & Use of Struct      & 28.68                                 & 38           \\ \hline
Library/SDK                             & Cargo              & 36.76                                 & 83           \\ \hline
Migration                               & Migration Problem  & 26.35                                 & 36           \\ \hline
\multirow{2}{*}{Parallelization}        & Mutex              & 33.63                                 & 71           \\
                                        & Parallel Execution & 34.2                                  & 80           \\ \hline
\end{tabularx}
\label{table:topic_difficulty_rust}
\end{table}

%% file: Tables/topic_corr.tex

\begin{table}[ht]
\centering
\caption{Correlation of topics popularity and difficulty.}
\begin{tabular}{|ccccc|}
\hline
\textbf{Language}                            & \textbf{\begin{tabular}[c]{@{}c@{}}Correlation\\ Coeff. /\\ p-value\end{tabular}}                                                                & \textbf{\begin{tabular}[c]{@{}c@{}}Avg. \\ Views\end{tabular}} & \textbf{\begin{tabular}[c]{@{}c@{}}Avg. \\ Score\end{tabular}} & \textbf{\begin{tabular}[c]{@{}c@{}}Avg. \\ Favourite\end{tabular}} \\ \hline
\multicolumn{1}{|c|}{\multirow{2}{*}{Swift}} & \begin{tabular}[c]{@{}c@{}}\% w/o Accepted\\ Answers\end{tabular}      & -0.515/0.029 & -0.503/0.034 & -0.401/0.099   \\
\multicolumn{1}{|c|}{}                       & \begin{tabular}[c]{@{}c@{}}Median Time\\ to Answer (min.)\end{tabular} & -0.381/0.119 & -0.244/0.33 & -0.096/0.705      \\ \hline
\multicolumn{1}{|c|}{\multirow{2}{*}{Go}}    & \begin{tabular}[c]{@{}c@{}}\% w/o Accepted\\ Answers\end{tabular}      & -0.72/0.006 & -0.764/0.002 & -0.659/0.014   \\
\multicolumn{1}{|c|}{}                       & \begin{tabular}[c]{@{}c@{}}Median Time\\ to Answer (min.)\end{tabular} & -0.476/0.1 & -0.575/0.04  & -0.448/0.124   \\ \hline
\multicolumn{1}{|c|}{\multirow{2}{*}{Rust}}  & \begin{tabular}[c]{@{}c@{}}\% w/o Accepted\\ Answers\end{tabular}      & -0.333/0.381  & -0.383/0.308 & -0.033/0.932   \\
\multicolumn{1}{|c|}{}                       & \begin{tabular}[c]{@{}c@{}}Median Time\\ to Answer (min.)\end{tabular} & -0.517/0.154  & -0.467/0.205 & 0.0/1.0      \\ \hline
\end{tabular}
\label{table: topic_correlation}
\end{table}

%% file: Sections/RQ4.tex
\subsection{RQ4. When were adequate resources available for the new programming languages in Stack Overflow?}
\label{RQ4}
\subsubsection{Motivation} 
The resources of a programming language, maturity and performance of its libraries usually take time to be stable. In the meantime, developers using that language are likely to discuss these in community QA sites such as SO. In the RQ, we would like to inspect the length time it takes for a language to get maturity by analyzing its footprint in SO.


\subsubsection{Approach} It is hard to define ``adequate resource'' of a programming language in a QA site. However, we can use an indirect approach to measure adequate resources. Two major types of Stack Overflow questions are \emph{repetitive} questions and \emph{new} questions. By \emph{repetitive} question we mean the same question or same problem was discussed before, but then the developers~\citep{TausczikWC17} faced it in another platform or environment. The decrease in the number of the new questions indicates that Stack Overflow already has the answer to most of the questions or problems. From this point of view, we can say that in Stack Overflow we have ``adequate resource'' of that particular language if the number of new questions is within a limit.

However, questions are not the only way developers interact with Stack Overflow. There are other ways like votes and comments. To consider all types of interactions into determining the expected time for the availability of adequate resources, we have followed the approach of Srba et al.~\citep{Srba2016}. Using this approach, we have calculated average post (by post we mean both question and answer) quality in Stack Overflow. To measure post quality, we need to consider all kinds of interactions within a given time frame. A deadline is needed to ensure that each post (both old and new) received equal time to receive votes and comments. Otherwise, old posts will get more time to get comments and votes than the new posts. In SO a post may receive a vote long after the date it was created. For example, in our dataset, a post received a vote from a user twelve years later. However, most votes, comments, and answers are available after a certain period. As stated by Bhat et al.~\citep{Bhat2014}, 63.5\% questions receive an answer within one hour, and only 9.98\% questions receive an answer after one day. To calculate post quality from votes of answers, accepted answers and comments, we have considered the votes received within thirteen days of the creation of the post. We studied the distribution of the accepted answer time, answer time, and comment time. We found that the \emph{thirteenth day} covers the 85 percentile of answer time, 95 percentile of accepted answer time, and 90 percentile of the comment time. We calculated the post quality increasing the duration, but found that quality did not change significantly. The \emph{quality score} represents the average post quality over a month and \emph{interaction score} represents the average developers' interaction of that language. To calculate \emph{quality score}, votes from accepted answers are given double weight compared to those without an accepted answer. This practice exists~\citep{Romano2013} to prioritize the contribution of accepted answers. The detail calculation of \emph{quality score} and \emph{interaction score} is presented below.

\noindent
Let, \\
$Q = $ {All questions of a month},\\
$A = $ {All answers of questions in $Q$ where creation date is within 13 days of  $Q$},\\ 
$C = $ {All comments of both $Q$ and $A$ within 13 days of $A$},\\
$S = $ {All accepted answers of $Q$ where creation date is withing 13 days of $Q$},\\
$T(x) = $ Creation time of item $x$.\\
Now,

\begin{equation}
\begin{split}
Interaction\ Score = \dfrac{\sum_{Q_i\in Q}Q_i+ \sum_{A_i\in A}A_i+\sum_{C_i\in C}C_i}{\sum_{Q_i\in Q}Q_i}
\end{split}
\label{eq: interaction score}
\end{equation}

\begin{equation}
\begin{split}
Quality\ Score = \dfrac{\sum_{Q_i\in Q}\sum_{\substack{Q_v\in Votes\: of\: Q_i\\T(Q_v) \leq T(Q_i)+13}}Q_v}{\sum_{Q_i\in Q}Q_i}+  \dfrac{\sum_{A_i\in A}\sum_{\substack{A_v\in Votes\: of\: A_i\\T(A_v) \leq T(A_i)+13}}A_v}{\sum_{Q_i\in Q}Q_i}\\ +\dfrac{\sum_{S_i\in S}\sum_{\substack{S_v\in Votes\: of\: S_i\\T(S_v) \leq T(S_i)+13}}S_v}{\sum_{Q_i\in Q}Q_i} 
\end{split}
\label{eq: quality score}
\end{equation}

\subsubsection{Results}
\begin{figure*}[t]
    \centering
    \subfloat[Post quality of new languages in Stack Overflow\label{fig:Content quality}]{{\includegraphics[scale=0.44]{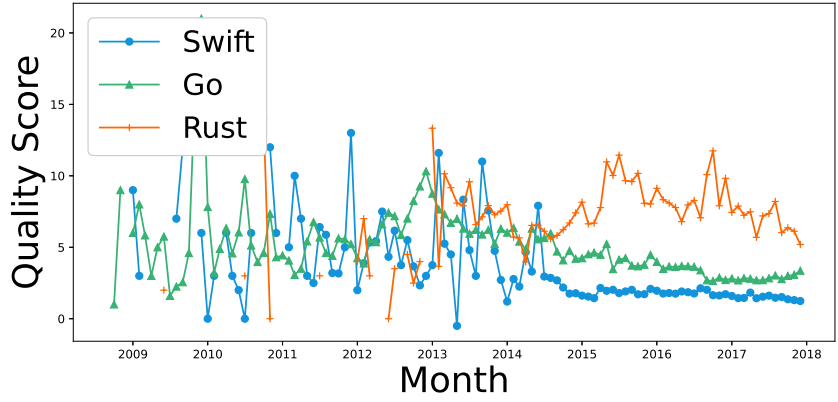} }}%
    \enskip
    \subfloat[Interaction of developers of new languages with Stack Overflow\label{fig:Interaction}]{{\includegraphics[scale=0.44]{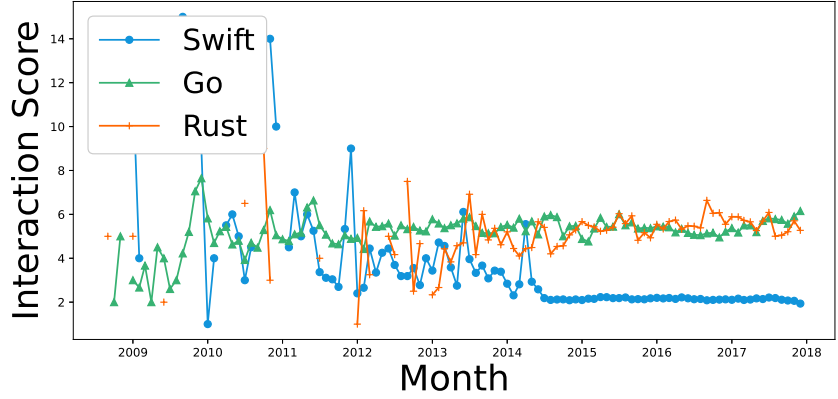}}}%
    \caption{Post quality and developers' interaction with new languages vs. time}%
    \label{score of languages}
\end{figure*}

We plotted the quality score and interaction score of the three languages in Figure \ref{fig:Content quality} and Figure \ref{fig:Interaction}, respectively. From Figure~\ref{fig:Content quality} it is quite clear that after the introduction of a language, post quality is unstable, and the quality scores are very high. The obvious reason behind this instability is that the language lacks resources, and every new release triggers a set of new questions. The questions of the starting years are less repetitive than the later years~\citep{Srba2016}, and it is the reason behind the high value of \emph{quality score}. Gradually \emph{quality score} stabilizes into a certain point. In a stable language, users' interaction with Stack Overflow should be minimum and within a range. From Figure~\ref{fig:Interaction}, it is evident that after the first release, \emph{interaction score} also stabilizes to a point which supports our conjecture.

\begin{figure*}[t]
    \centering
    \subfloat[Swift]{{\includegraphics[scale=0.25]{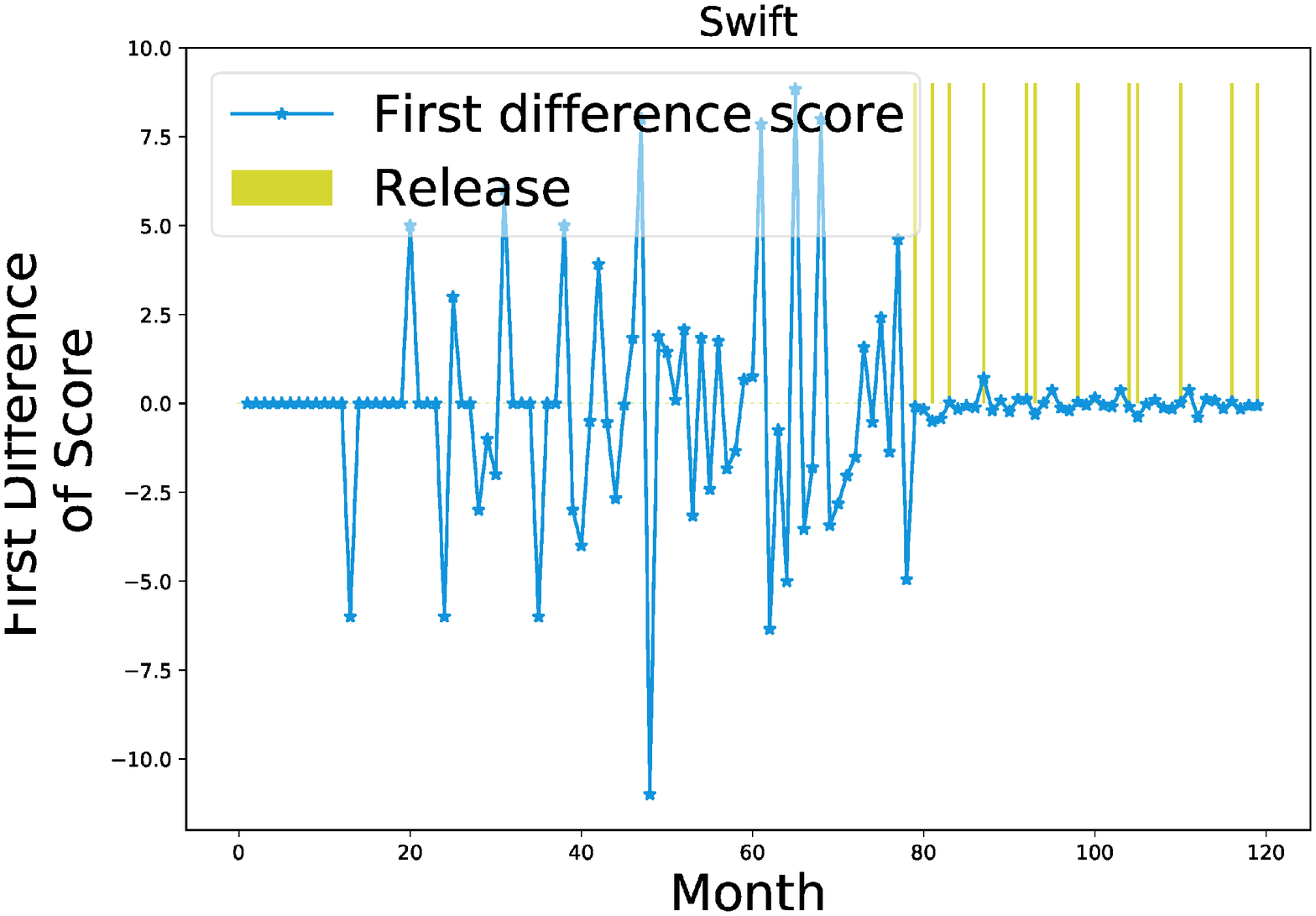} }}%
    \subfloat[Go]{{\includegraphics[scale=0.25]{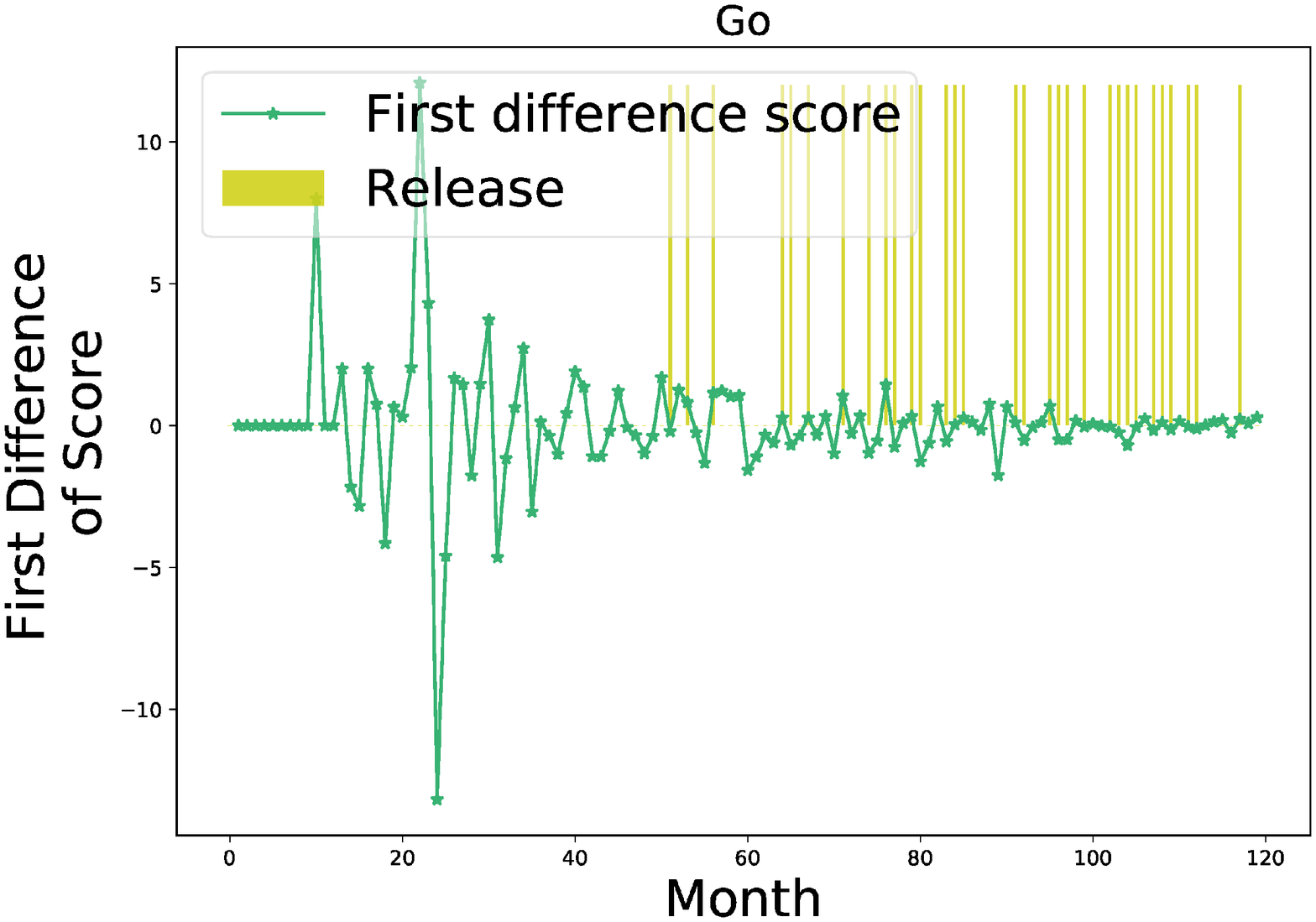}}}%
    \enskip
    
    \subfloat[Rust]{{\includegraphics[scale=0.25]{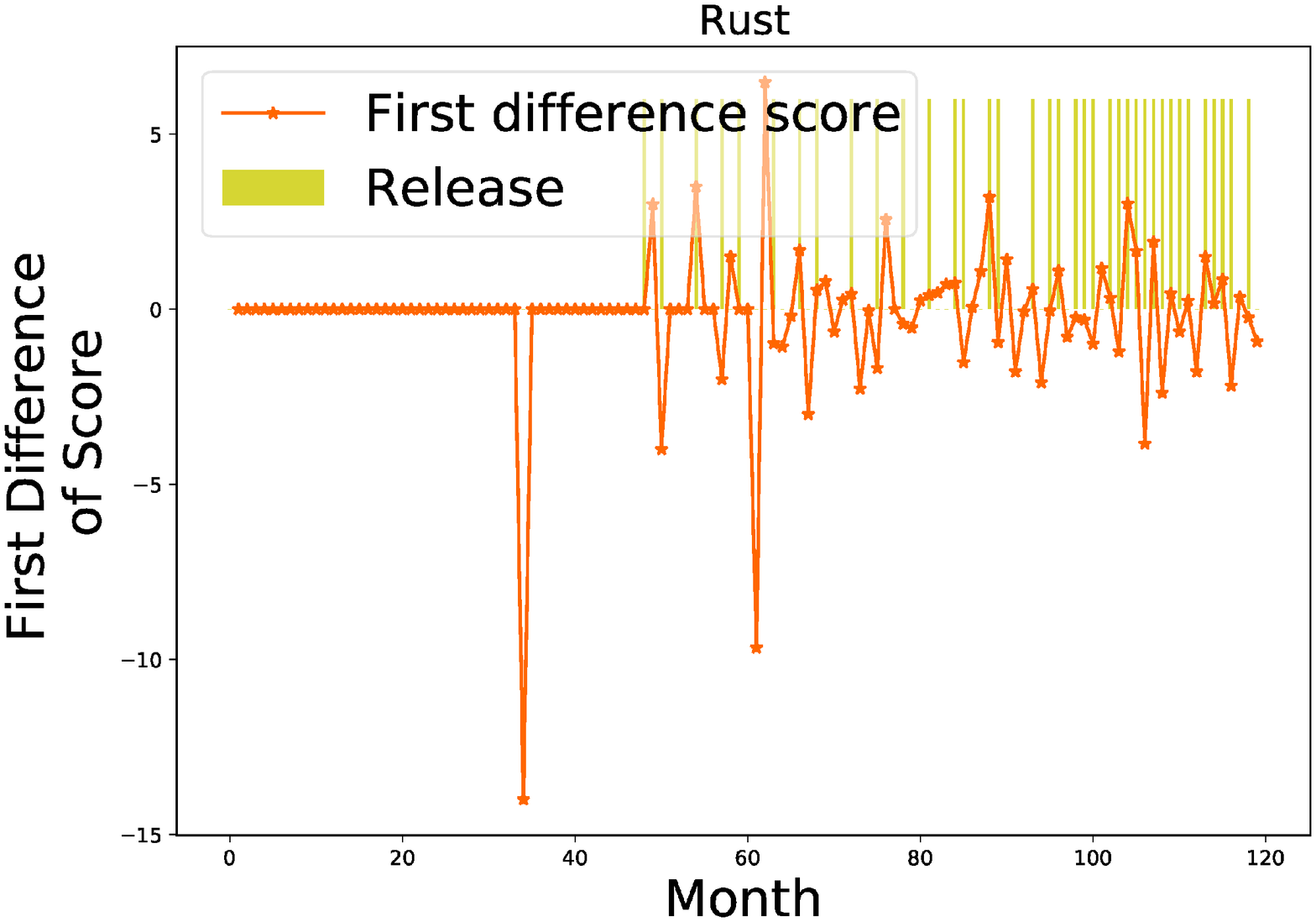}}}%
    \caption{First difference of the post quality and release of a new version of new languages.}%
    \label{First difference and release}
\end{figure*}

To effectively measure the difference of quality scores between consecutive months, \emph{first difference} metric~\citep{Rasheed2011} has been applied to the quality score of each language. The first difference of quality score is the difference of quality scores between two consecutive months. The \emph{first difference} technique removes any unobserved variable from data. Moreover, as the data points are taken at a constant interval, the value of the \emph{first difference} works like a differential value of the quality score function from where we can observe the change. The first difference is plotted against the release time in Figure~\ref{First difference and release}. In the beginning, the first difference of quality score was following the trend of release. However, gradually it decreases the level of response which means the language is stabilizing. We can detect a stable point for a language from this point of view. By the stable point, we mean the starting date of the period after the first release of a language when the language is so stable that a single release cannot change or disrupt the development process.

If the first difference of quality score of a language is within a range, then it has two implications. First, the language is stable and it does not initiate any significant change in the development lifecycle. Secondly, most of the Stack Overflow posts are repeating for this language and the contribution of these kind of questions will be omitted in the first difference process. Now, we have the effect of the change of frequency of new questions in the first difference of quality score. Therefore, the first difference of quality score within a range means developers face fewer problems that are not already answered in Stack Overflow. Hence, we can say at this point that the new languages have adequate resources in Stack Overflow.


We defined the stable point as the time point after which the value of the first difference is always between -1 and 1. Stable points for each language are presented in Table~\ref{table:Stable point}

\begin{table}[htbp]
\centering
\caption{Stable point of the new languages}
\begin{tabular}{|l|l|l|}
\hline
\textbf{Language} & \textbf{Release Date} & \textbf{Stable Point Date}  \\ \hline
Go & March 1, 2012 & July 1, 2015\\ \hline
Swift & September 9, 2014 & November 1, 2016\\ \hline
Rust & January 1, 2012 & Not reached \\ \hline
\end{tabular}
\label{table:Stable point}
\end{table}

In Stack Overflow, the number of Rust developer is too low compared to the other two languages. It is quite common in Stack Overflow that a particular portion of developers leaves or becomes inactive in Stack overflow after some time. The post quality of Rust will change quickly after such departure. However, such departure cannot change Go or Swift post quality so frequently as departing developers represent a small percentage of the whole community of these languages in Stack Overflow. We also observed that the Rust language's release frequency is relatively high compared to the other two languages. These can be reasons why Rust has not reached the stable point yet.

\boxtext{In Stack Overflow, we can expect adequate resources for Swift after two years of release, while this period is three years for Go. We have found the evidence of having an inadequate resource of Rust language in Stack Overflow.}

\boxtext{The size of an active community can influence the growth of a new language.}


%% file: Sections/RQ7.tex
\subsection{RQ5. Is there any relationship between the growth of the three programming languages and developers' activity patterns?}
\label{RQ7}

\subsubsection{Motivation} 
Along with the QA sites such as Stack Overflow, repository like GitHub presents the activities of the developers on that language. In this research question, we would like to explore these two sources to understand how language advancement is reflected in the developer activities and engagements in both these platforms.


\subsubsection{Approach} The most common developers' activities in SO~\citep{Badashian2014} are:
\begin{enumerate}[leftmargin=10pt, itemsep=0pt]
    \item \textbf{Questions:} Developers ask development-related questions. Questions might be moderated based on clarity and duplicity.
    \item \textbf{Answers:}  Developers answer questions about their field of expertise.
    \item \textbf{Comment:} Users can comment on other users' questions and answers.
    \item \textbf{Up Votes:} Developers can vote to increase the score of other users' questions or answers.
    \item \textbf{Down Votes:} Developers can cast votes to decrease the score of other users' questions or answers.
    \item \textbf{Question View:} Users can view other users' questions. (SO does not keep this count with a timestamp).
    \item \textbf{Answer View:} Users can view other users' answers. (SO does not keep this count with a timestamp).
\end{enumerate}
High developers' activity helps to expose special cases and rare bugs of a project. Developers use the issue to inform the language owners about these problems or a particular case. The solution to these problems and bugs led to the growth of the language.
Hence, we expect a relationship between the issue and the developer's activity pattern. Moreover, developers' activity can also be observed from the number of users and repositories of that language from GitHub. Table ~\ref{table:model parameters} summarizes the descriptions of and rationales behind the studied factors. To measure the relationship among variables, we performed the following steps.
\input{Tables/Model_parameters}
\begin{enumerate}
\item Model Construction (MC)
\item Model Analysis (MA)
\end{enumerate}
These steps are discussed below.

\noindent\textbf{Model construction (MC):} 
We build a regression model to explain the relationship between dependent and explanatory variables. The regression model fits the dependent variable with respect to independent variables. We followed the model construction approach of Harrel et al.~\citep{Harrell2015}. While relaxing the linearity assumption, this approach models the nonlinear relationship accurately.
The steps for model construction are described below.
\begin{enumerate}[leftmargin=10pt]
\item \emph{Estimation of maximum degrees of freedom:}
A critical concern in model building is overfitting. Overfitting is most frequent in models that use more degree of freedom than the dataset can support. Hence, we have fixed the maximum degree of freedom for our model. As suggested by Harrel et al.~\citep{Harrell2015}, we have fixed \(\frac{n}{15} \) degree of freedom for our model, where n is the number of data points (120) in the dataset.
\item \emph{Normality adjustment:}
We fit our regression models using the Ordinary Least Squares (OLS) technique. OLS assumes the normality in the distribution of the dependent variable. Hence, it is crucial that the distribution of the dependent variable is normal. A widely used approach for conversion into a normal distribution is applying $\ln$ function~\citep{pmid25092958}. We have some zero value in our dataset. Therefore in our case, we have used $\ln (x+1) $ to lessen the skew and better fit the OLS assumption.
\item \emph{Correlation analysis:}
Before building the model, we checked the highly correlated explanatory variables. In this step, we have used Spearmen rank correlation as it is resilient to data that is not normally distributed. We have constructed a hierarchical overview of the correlation among explanatory variables. For sub-hierarchies of explanatory variables with correlation $\rho$ \textgreater 0.9, we selected only one element of the sub-hierarchy.
\item \emph{Fit regression Model:}
Finally, after selecting explanatory variables and log transformations of dependent variables, we fit our regression models to the data.
\end{enumerate}

\noindent\textbf{Model Analysis (MA):} We have calculated the adjusted $R^2$ to measure the goodness of fit of the model. Adjusted $R^2$ considers the bias of an additional degree of freedom by penalizing the model for each degree of freedom.
The steps for model analysis are described below.
\begin{enumerate}[leftmargin=10pt]
\item \emph{Assessment of Model stability:} 
Adjusted $R^2$ may overestimate the performance of the model for the curse of overfitting. The performance estimation is taken into account by subtracting the average \emph{optimism}~\citep{Efron1986}. The \emph{optimism} is calculated in three steps. First, a bootstrap is created from N samples. Second, a model is fitted on the bootstrap data using the same degree of freedom. Third, \emph{optimism}, the difference between the adjusted $R^2$ of the bootstrap model and the model built in the previous step(original model) is calculated. The process is repeated for 1000 times, and we have got average optimism. Finally, we subtracted the average optimism from the original adjusted $R^2$ and got optimism reduced $R^2$.
\item \emph{Estimation of the power of explanatory variables:}
To measure the impact of an explanatory variable on a model, we measured the difference in performance between all explanatory variables (full model) and all explanatory variables except one (dropped model). A $\chi^2$ test is applied to the resulting values to detect whether each explanatory variable improves model performance to a statistically significant degree. To estimate the impact, we performed the Wald $\chi^2$ maximum likelihood test. The larger the Wald $\chi^2$ value, the more significant the impact of that particular explanatory variable is~\citep{McIntosh2015}.
\end{enumerate}

We can observe the relation between developers' activity patterns and advancement of the language project from two different perspectives: (1) Question count of Stack Overflow and (2) Repository and User count of that language from GitHub. Hence, we performed the process of estimating the relationship for each perspective. 

We have used open issue count, closed issue count, and the ratio of open issue count with respect to the total number of the issue as the explanatory variable and the question count as the dependent variable to estimate the relationship between issue frequency and developers' activity from Stack Overflow perspective. We collected 47710, 23967, 14033 issue data for Rust, Go, and Swift language, respectively.

Developers use an issue to ask owners about a new feature and seeking help with any problems. More developers may lead to a high number of issues. Therefore, from the GitHub perspective, to model a relationship between the issue and the developers' activity, we have used the User count and the Repository count of that language as the explanatory variable and Open issue count as the dependent variable. We have collected the number of repositories and users for each language. To compute the number of new users for each language, we searched for all the users whose account creation date is within a particular month and whose major language is this language.

\subsubsection{Results}
According to our approach, results for the estimation of the relationship between the age of the language and the developers' activity from the \textbf{perspective of Stack Overflow} is presented below.
\input{Tables/issue_question_relation.tex}

\begin{enumerate}[wide=0pt, leftmargin=10pt]
    \item[(MC-1)] \textbf{Estimation of the maximum degree of freedom:} We have 120  points in our dataset. Hence, by Harrel et al.~\citep{Harrell2015} we can allow maximum 8 degrees of freedom.
    \item[(MC-2)] \textbf{Normality adjustment:} Question frequency of the new languages is right-skewed. So we have to perform log normality adjustment in this case.
    \item[(MC-3)] \textbf{Correlation analysis:}
    We hierarchically clustered the features by Spearmen $\mid \rho \mid$ value. It is found that the open issue ratio is highly correlated with closed issue count. For the sake of completeness, we have created a model using closed issue count instead of open issue ratio and vice-versa but have not found any change in the performance of the model.
    \item[(MA-1)] \textbf{Assessment of model stability:}
    Table~\ref{table:issue_question relationship} presents the adjusted $R^2$ and optimism corrected $R^2$. From Table~\ref{table:issue_question relationship}, we can say that the model is stable for Swift and Rust where the optimism (the difference between Adjusted $R^2$ and Optimism-reduced $R^2$) is 0.002 and 0.005 respectively. However, for Go, the optimism is 0.011. Though the difference is noteworthy, it does not invalidate our model.
    \item[(MA-2)] \textbf{Estimation of the power of explanatory variables:}
    The high $\chi^2$ value of the open issue and the ratio of the open issue to the total number of issue in  Table~\ref{table:issue_question relationship} represents the significant role of these parameters in Stack Overflow. However, they are not that much significant for determining the number of Swift and Rust language questions in Stack Overflow. On the other hand, closed issue ratio is significant for all three languages in determining the number of questions in Stack Overflow which is proved by the high $\chi^2$  value of closed issue in Table~\ref{table:issue_question relationship}. Overall, the $\chi^2$ value of Swift language is relatively smaller than the other two languages. Hence, we can say that the GitHub issue provides a meaningful and robust amount of explanatory power in describing question frequency of new languages except Swift.
\end{enumerate}
It is quite clear from the adjusted $R^2$ value of Table~\ref{table:issue_question relationship} that there is a relationship between the growth of a language and the number of questions posted. As seen from Table~\ref{table:issue_question relationship}, the number of closed issues is the most impactful explanatory variable for the Rust language model. Hence, we can say that the number of open issues will significantly influence the number of Rust question in Stack Overflow.

The relationship between the growth of a language and the developers' activity from the \textbf{perspective of GitHub} is presented below.
\input{Tables/github_question_relation.tex}

\begin{enumerate}[wide=0pt, leftmargin=*]
\item[(MC-1)] \textbf{Estimation of the maximum degree of freedom:} To answer this question we have used the same dataset used in the previous step. Hence, we can allow a maximum of 8 degrees of freedom.
\item[(MC-2)] \textbf{Normality adjustment:} Like the previous step, we have applied a log transform to normalize the dependent variable (open issue count).
\item[(MC-3)] \textbf{Correlation analysis:}
 We have used two features to build this model. Hence, instead of hierarchical clustering, we just calculated the Spearmen $\mid \rho \mid$ value between \emph{user count} and \emph{repository count}. It is found that they are not correlated.
\item[(MA-1)] \textbf{Assessment of model stability:}
Table~\ref{table:github_question relationship} presents the adjusted $R^2$ and optimism reduced $R^2$. From Table~\ref{table:github_question relationship} the optimism for each language is \textless 0.01 which ensures the stability of the model.
\item[(MA-2)] \textbf{Estimation of the power of explanatory variables:}
The issue is associated with the developers' experience. Hence, we expect the `User' parameter to be an important feature in determining the number of open issue in the official GitHub repository of new languages. Table~\ref{table:github_question relationship} shows a high $\chi^2$ value of user count parameter, which supports our conjecture about the significance of the number of users in determining the number of open issue in GitHub. We have also found that Rust has relatively fewer users in GitHub than the other two languages which are expressed in the $\chi^2$ value of the `User' parameter for Rust. The high $\chi^2$ value of the repository parameter for Swift and Rust language represents the significance of the number of the repository in determining the number of Swift and Rust open issue. However, the number of repositories is less significant in determining the number of open issue in the Go GitHub repository than the other two languages.
\end{enumerate}
From the adjusted $R^2$ value of Table~\ref{table:github_question relationship}, it is clear that there is a strong relationship between developers' GitHub activity and the number of open issue in the official repository of that respective language.

\boxtext{There is a relationship between developers' activity pattern and the growth of the language.}

\boxtext{The number of open issues of Rust in GitHub significantly influenced the number of questions on Rust in Stack Overflow.}

\boxtext{The open issue count of Swift and Rust is highly dependent on the number of repositories of those languages in GitHub.}

Every new release impacts the growth of a programming language. After a new release, the developers' activity can give us an idea about the relationship between developers' activity pattern and the growth of a language. To observe the developers' activity pattern after a new release, we have collected all release dates of new languages from GitHub and then plotted them alongside question, issue, and repository count.

\begin{figure*}[t]
    \centering
    \subfloat[Swift]{{\includegraphics[scale=0.15]{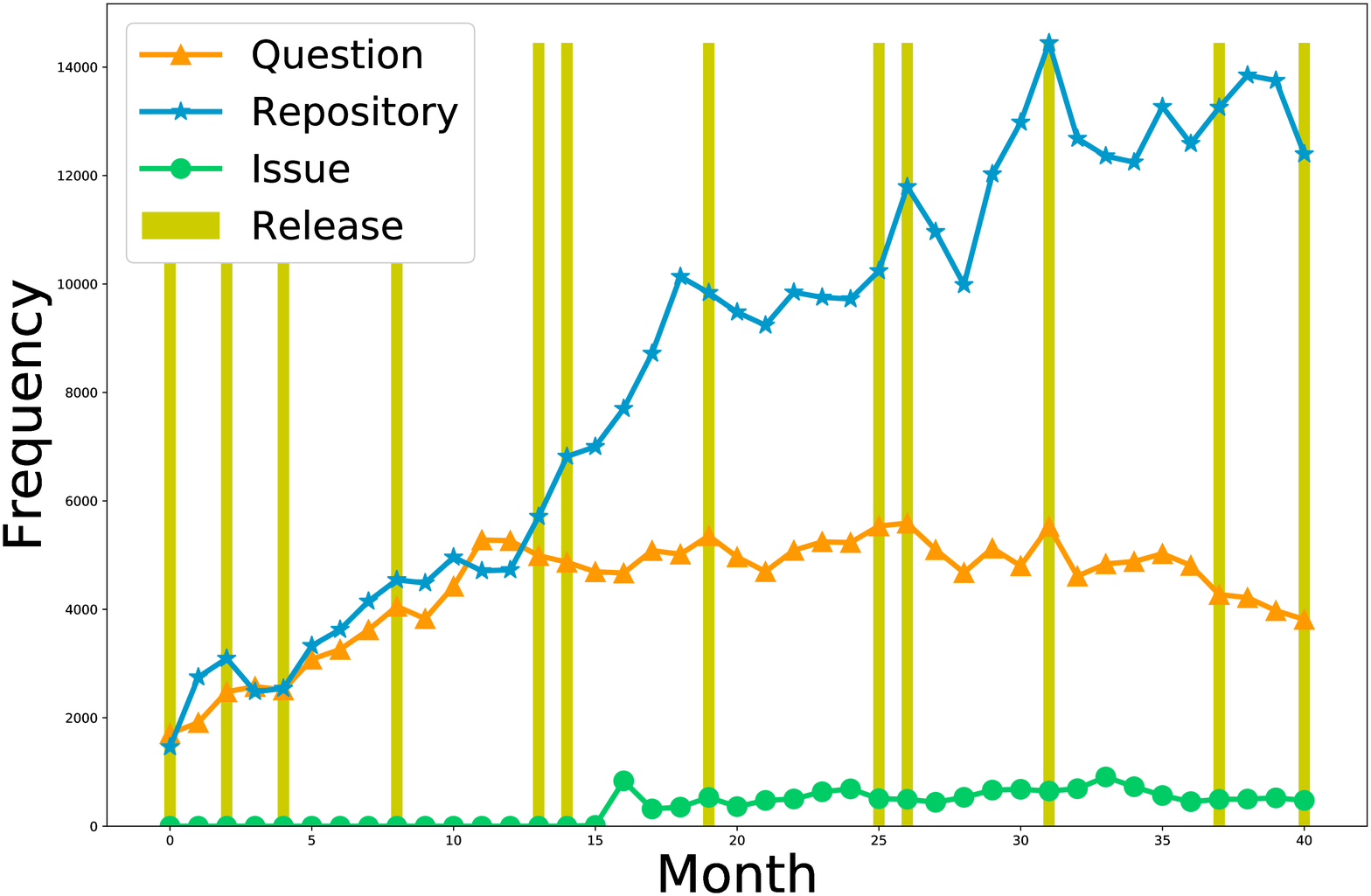} }}%
    \enskip
    \subfloat[Go]{{\includegraphics[scale=0.15]{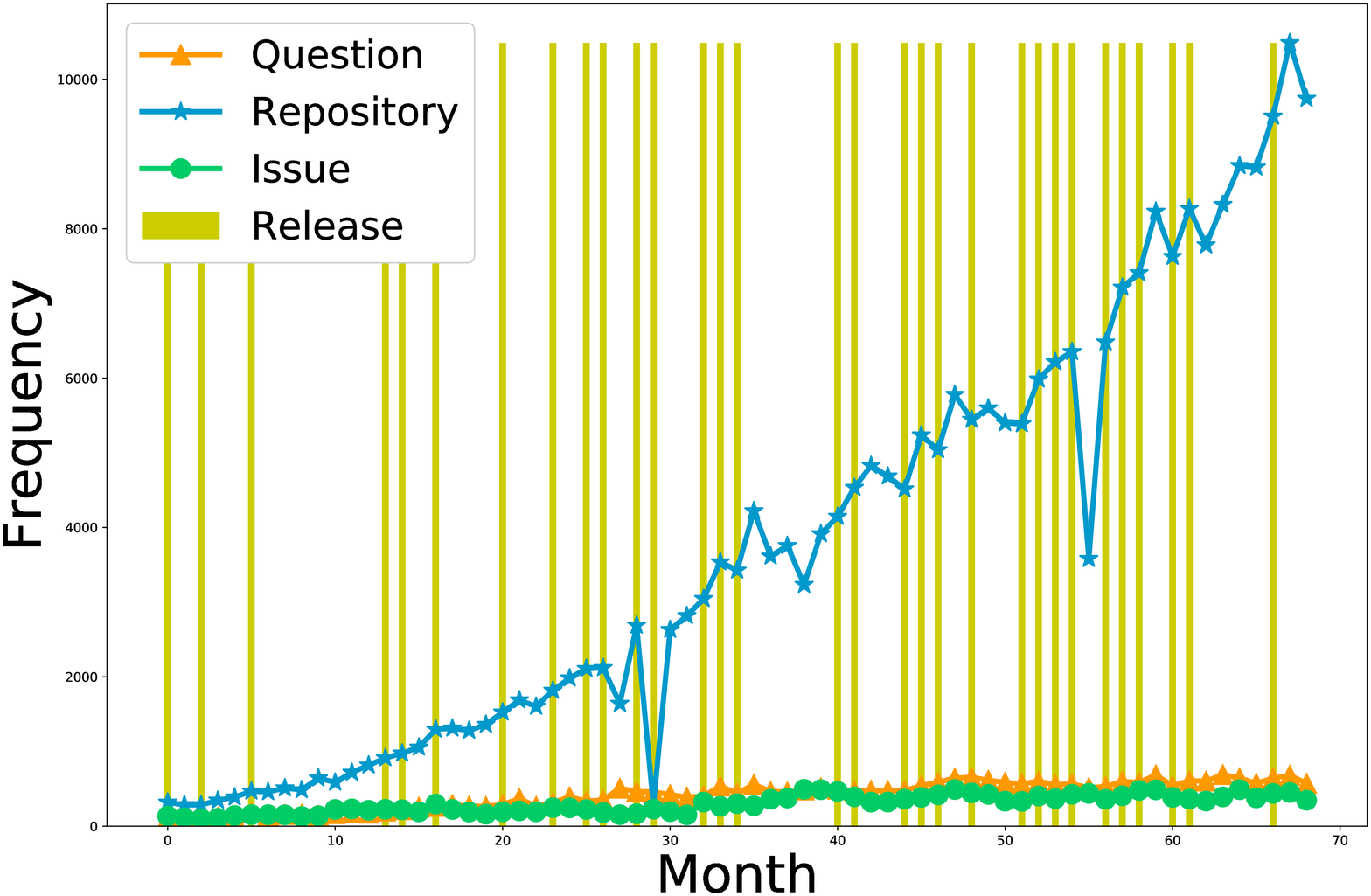}}}%
    \enskip
    \subfloat[Rust]{{\includegraphics[scale=0.15]{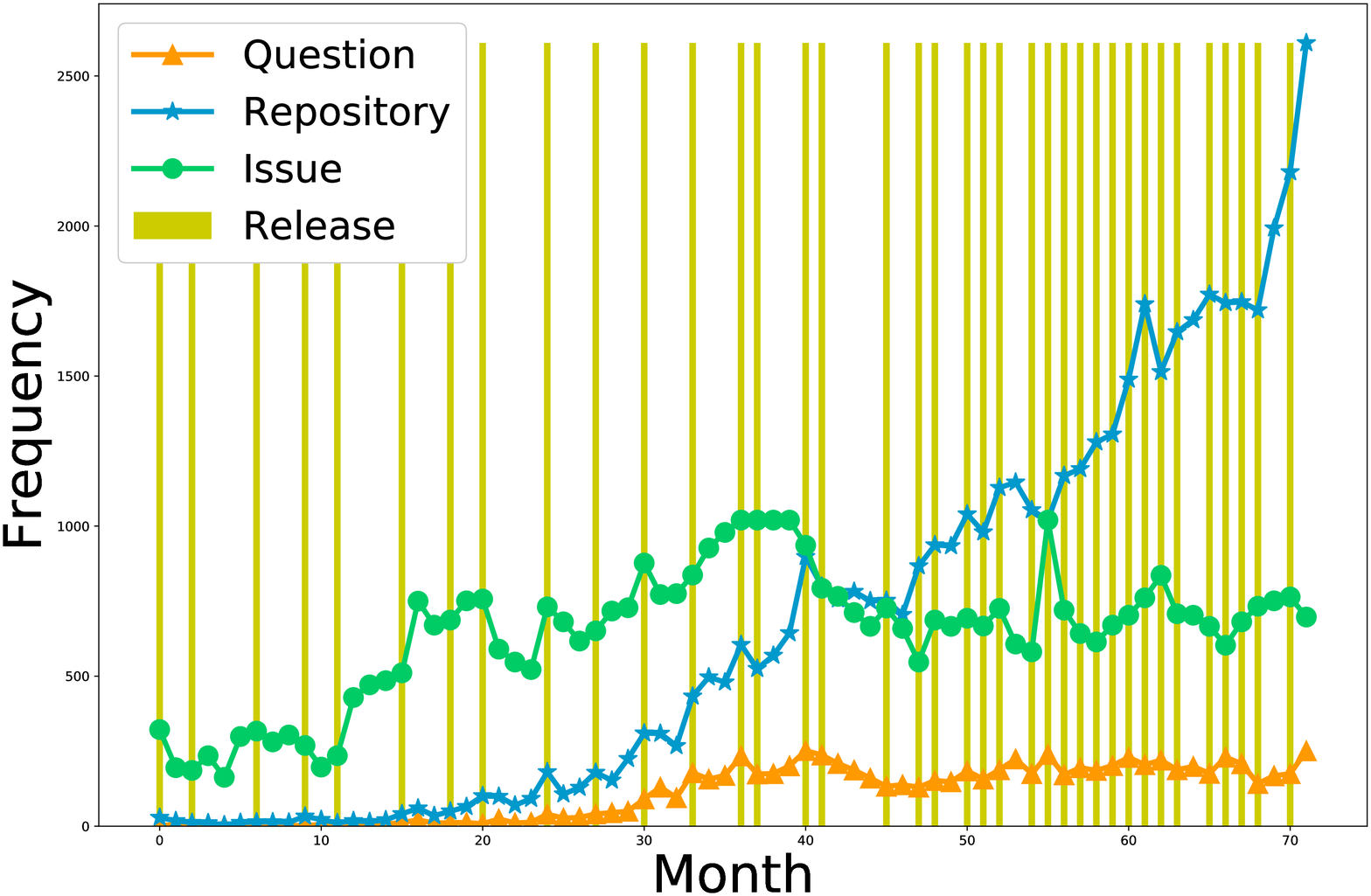}}}%
    \caption{Release of a new version and developer activity pattern per languages}%
    \label{fig:Release and user behavior}%
\end{figure*}

Each sub-figure shows the response after the release of a new version. It is clear that the developers' activity is influenced by the benefits, features, and bugs of the new release. This trend is also visible in question count, i.e., question count increases after each release (Figure~\ref{fig:Release and user behavior}). However, we observed that the issue count of Swift is less influenced than that of the other two languages. For tracking bugs and language problems, \href{https://Golang.org/project/}{Go} and \href{https://github.com/Rust-lang/Rust/blob/master/CONTRIBUTING.md}{Rust} use only GitHub issue tracker. On the other hand, besides using the GitHub issue tracker, \href{https://Swift.org/contributing/\#reporting-bugs}{Swift} uses their own JIRA~\citep{wiki:JIRA} instance for \href{https://bugreport.apple.com/}{tracking bugs}. This is a likely cause of the difference, and as a result, the issue count of Swift represents a portion of the actual issue (bugs), so it is less influenced by a new release than the other two languages. We have tested this hypothesis using statistical testing. We have performed Wilcoxon signed-rank test between the question, repository, and issue count of the month before release and the count of the month of release. The result is presented in Table~\ref{table:release_impact}. However, only the change in repository count was significant. The reason behind this significance is after each new release developers create a new repository to test the new features without altering the production version of the software. Hence, the number of repository increases after a new release. Though we can observe that the question and issue count 
\input{Tables/release_impact.tex} are responding with the release of a new version, it is statistically insignificant according to the Wilcoxon signed-rank test. To find the reason behind this insignificance, we conducted a further investigation. We noticed that the spikes in question and issue count curve does not appear immediately after a release. Rather, it appears after a variable time gap. Thus, the change in question and issue count is statistically insignificant.

%% file: Tables/Model_parameters.tex
\begin{table}[!h]
\centering
\caption{The description and rationale for the factors used in the regression model}
\begin{tabular}{lll}
\hline
\textbf{Factor name}                & \textbf{Description} & \textbf{Rationale} \\ \hline

 \multicolumn{3}{c}{Factors based on the number of questions posted in Stack Overflow}           \\ \hline

Open Issue Count           & \begin{tabular}[c]{@{}l@{}}The number of open \\issue in the official\\ repository of that\\ language\end{tabular}           &    \begin{tabular}[c]{@{}l@{}}This factor reflects\\the maturity of\\language\end{tabular}       \\ 

Closed Issue Count         &   \begin{tabular}[c]{@{}l@{}}The number of closed \\issue in the official\\ repository of that\\ language\end{tabular}          &    \begin{tabular}[c]{@{}l@{}}This factor reflects\\the maturity of\\language\end{tabular}       \\ 

Ratio of Open Issue        & \begin{tabular}[c]{@{}l@{}}The ratio of the  open \\and closed issue in\\ the official repository\end{tabular}             &    \begin{tabular}[c]{@{}l@{}}This factor reflects\\the agility of\\the language project\end{tabular}        \\ \hline

\multicolumn{3}{c}{Factors based on the Number of open issue in GitHub}                   \\ \hline

User Count                 &  \begin{tabular}[c]{@{}l@{}}The number of new \\users of this language\\ on GitHub\end{tabular}            &   \begin{tabular}[c]{@{}l@{}}This factor reflects\\the effect of incomplete\\ release over the\\ reputation of the language\end{tabular}        \\ 

Repository Count           &  \begin{tabular}[c]{@{}l@{}}The number of new\\repositories of this\\language\\ on GitHub\end{tabular}        &     \begin{tabular}[c]{@{}l@{}}This factor reflects\\the effect of incomplete \\release over the\\ developer base\end{tabular}      \\ \hline
\end{tabular}`
\label{table:model parameters}
\end{table}

%% file: Tables/issue_question_relation.tex
\begin{table}
\centering

\caption{Coverage model statistics based on Stack Overflow (* p\textless 0.1 , ** p\textless 0.001)}
\begin{threeparttable}
\begin{tabular}{|l|l|l|l|}
\hline
       & \textbf{Swift} & \textbf{Go} & \textbf{Rust} \\ \hline
    Adjusted $R^2$   &0.443  & 0.813 &0.924  \\ \hline
    Optimism-reduced $R^2$ & 0.441 &0.802  &0.919 \\ \hline
    Budgeted Degree of freedom & 8 & 8  &8 \\ \hline
    Open Issue $\chi^2$             & 10 \tnote{*} & 53\tnote{**} & 10\tnote{*}\\ \hline
    Closed Issue $\chi^2$           & 59 \tnote{*}& 251\tnote{**} & 716\tnote{**} \\ \hline
    Ratio of Open Issue $\chi^2$    & 10 \tnote{*} & 63\tnote{**} & 21\tnote{**}\\ \hline
\end{tabular}

\end{threeparttable}
\label{table:issue_question relationship}
\end{table}

%% file: Tables/github_question_relation.tex
\begin{table}
\centering
\caption{Coverage model statistics for relationship based on GitHub (* p\textless 0.1 , ** p\textless 0.001)}
\begin{threeparttable}
\begin{tabular}{|l|l|l|l|}
\hline
       & \textbf{Swift} & \textbf{Go} & \textbf{Rust} \\ \hline
    Adjusted $R^2$   &0.755  & 0.907 &0.862 \\ \hline
    Optimism-reduced $R^2$ & 0.751 &0.905  &0.859 \\ \hline
    Budgeted Degree of freedom & 8 & 8  &8 \\ \hline
    User $\chi^2$             & 64 \tnote{*} & 215\tnote{*} & 50\tnote{*}\\ \hline
    Repository $\chi^2$           & 289 \tnote{**}& 19\tnote{**} & 192\tnote{**} \\ \hline
\end{tabular}
\end{threeparttable}
\label{table:github_question relationship}
\end{table}

%% file: Tables/release_impact.tex
\begin{table}
\centering
\caption{Wilcoxon signed rank test result for the comparison of change between before and after of a release.}
\begin{tabular}{|c|c|c|c|}
\hline
\multicolumn{1}{|l|}{\textbf{Language}} & \textbf{\begin{tabular}[c]{@{}c@{}}Question\\ p value\end{tabular}} & \textbf{\begin{tabular}[c]{@{}c@{}}Repository\\ p value\end{tabular}} & \textbf{\begin{tabular}[c]{@{}c@{}}Issue\\ p value\end{tabular}} \\ \hline
Swift                          & 0.583                                                      & \textless{}0.01                                              & 0.6                                                     \\ \hline
Go                             & 0.149                                                      & \textless{}0.01                                              & 0.433                                                   \\ \hline
Rust                           & 0.473                                                      & \textless{}0.01                                              & 0.2                                                     \\ \hline
\end{tabular}
\label{table:release_impact}
\end{table}

%% file: Sections/Implication.tex
\section{Implication}
\label{sec:Implication}
Thus far, we have discussed the characteristics of answer pattern of new languages, the relation between the advancement of new languages and its developers' activity, and expected answer interval for new languages. In this section, we discuss the implications of our findings. As well as helping developers find resources while learning a new language, our study can also help language owners, researchers, and Stack Overflow to refine their strategies to support the growth of new languages. 

\indent \textbf{Developers:} In this study, we have estimated the answer interval and time when we can expect the availability of the adequate resource in Stack Overflow. If the community support is still evolving in Stack Overflow, developers can decide to look into other resources. Sometimes community projects developed and curated by developers can be an alternative for traditional resources. For example, Rust was a community project of concerned developers. After strong positive feedback, it was donated and has been part of official Rust documentation from Rust 1.25.

\indent \textbf{Language owners:} Our study identifies the significant difference in answer interval between two phases of new languages. As support for the developers in the starting stages is likely to play a significant role in the overall acceptance of that language, owners should provide extensive support during that time. Another option for new languages that are currently in the design stage can be to use the community base of some mature language by carefully selecting predecessor language. Moreover, new languages can fill the gap in supporting materials using developer-friendly documentation with a detailed example. We observed that the issue and release version influences developers' activity pattern (Table~\ref{table:issue_question relationship}, Table~\ref{table:github_question relationship}, Figure~\ref{fig:Release and user behavior}). Though it is not possible to release a bug-free version, extra care must be taken for a bug-free release and solution of issues in GitHub. A good portion of questions in Stack Overflow seeks clarification of the documentation. Owners should take extra care to prepare documentation suitable for developers of all levels. We have also found that migration is a common topic among all the new languages. As there are many mature languages in the same domain before the arrival of new languages it is assumed that a huge number of new language projects are migrated projects from some other language. To facilitate developers' efforts, language owners should provide detailed documentation of migration steps from common sources.

\indent \textbf{Stack Overflow:} Small community size can disrupt the growth of a language. Our study found that the new languages have a small number of experts or active developers in Stack Overflow. To support the growth of a language which has a few expert developers, Stack Overflow should refine their strategy. According to the current policy, the stack overflow focuses on the expert developers. However, to support new languages, they should encourage developers from all levels to answer questions. It supports the findings of Srba et al.~\citep{Srba2016} where they suggested Stack Overflow replace the current question-oriented policy with an answer-oriented policy.

\indent \textbf{Researchers:}  We have found that migration is the hardest topic in two of the three new languages in terms of posts without an accepted answer. Furthermore, it is a common topic in all of the three languages. As migration problems are too user-specific further research may be conducted on how a generalized solution can be designed to  solve user-specific issues. The data and data structure category are common and one of the top two categories in all the languages in terms of the number of posts. It gives the researchers a direction of an impactful and broad research area. Our study finds that the library/SDK category is a common discussion topic among new language developers. Also, this category contains one of the top three difficult topics in all three languages. We have observed that developers often face difficulties integrating libraries or setting up communication between SDKs'. A standard protocol for SDK communication may help developers to overcome such difficulties.


%% file: Sections/Validity.tex
\section{Threats to validity}
\label{sec:validity}
In this section, we discuss the validity of our study.

\indent \textbf{Internal validity:} Use of tags to categorize questions by language is an internal threat to validity. A new Stack Overflow user may not add an appropriate tag with the question. However, Stack Overflow questions go through an extensive moderation process, and eventually, it will have the appropriate tags. In some cases, our identification of posts by tags may not capture the posts of new languages. To alleviate this threat, we considered the relevance of tags. In this study, we have used Stack Overflow as the primary dataset. There are many language-specific developers' forums and QA sites, and those sites may contain posts that can help to understand the growth of new languages. However, we believe that numerous participants and the widespread popularity of Stack Overflow have made it a familiar venue for developers. Hence, the posts of Stack Overflow are considered enough to understand the trends of the growth of a new language.

\ra{We conducted this study with the Stack Overflow (SO) dump of January 2018, which was the latest dump available during our analysis. Our analysis presents the types of discussions and support developers offer in SO regarding the three new programming languages Go, Swift, and Rust. While these three languages are new compared to languages like Java and C\#, we note that we found at least three years of data for each language (Go, Swift, Rust) in SO in our dump of January 2018. Such a large volume of data can provide us considerable insights into the research questions we explored in our paper. However, the data dump is a little bit old, and replicating this study on a newer dataset may lead to different results. Like any language, over time, a new language is no longer considered new. This can be true for the above three new languages if they are studied for a longer period. Given that our focus was to understand how these three new languages are discussed and supported, our analysis and results from January 2018 across the three languages are well-suited to accommodate the analysis of newness/freshness of the three languages.}

\indent \textbf{External validity:} After the inception of the Stack Overflow (2008), about 35 programming languages have been released~\citep{wiki:Timeline}, whereas this study is focused on three languages (Swift, Go, and Rust). For this reason, our research results may not apply to other new languages. However, in this study, we did not emphasize any specific feature of a particular language. The languages we considered vary in terms of their time of inception and other properties (such as having predecessor language or not). Instead, we focused on the characteristics and trends of the growth of new languages. We compared the growth trends with a top-tier (Java) and mid-tier (Python) language and found that mid-tier language (Python) shows similar characteristics that confirm the generalizability of the findings. The dissimilarity with the top-tier language (Java) is that we missed the community interaction at the initial period of this language. Java was published a long time ago and was already a developed language before the establishment of SO. Therefore, we think our findings are free from any bias in a particular language.

%% file: Sections/RelatedWorks.tex
\section{Related Works}
\label{sec:Related Works}

There have been many works on Stack Overflow data, analyzing the developers' discussion topics. Barua et al.~\citep{Barua2012} have investigated the question "What are the developers asking?". Rosen and Shihab~\cite{Rosen2015}, Bajaj et al.~\citep{Bajaj2014}, Wan et al.~\cite{Wan2019}  did similar work focusing on mobile developers, web developers, and blockchain developers, respectively.

In a study on big data developers, Bagherzadeh et al.~\cite{Bagherzadeh2019} have identified no statistically significant correlation between big data topics' popularity and difficulty. To reach this conclusion, they used LDA to identify big data topics and then calculated the topics' popularity and difficulty. However, in a similar study on concurrency developers, Ahmed et al.~\cite{Ahmed2018} have found a negative correlation between difficulty and topics' popularity.

Abdellatif et al.~\cite{Abdellatif2020} conducted a study on chatbot developers to identify challenging chatbot development issues. For this study, they extracted posts from stack overflows related to chatbot development. They found that the maturity level of the chatbot community is still lower than other SE fields. One of their suggestions for facilitating chatbot developers' efforts is the improvement of the documentation of the chatbot platform and documentation for integration with popular third parties.

Hart and Sharma~\citep{Hart2014} had suggested to consider user reputation, the social reputation of the answerer, and post length to judge post quality.

Reboucas et al.~\citep{Reboucas2016} have compared the data from Stack Overflow with opinions of 12 Swift developers to answer three research questions - common problems faced by Swift developers, problems faced by developers in the usage of `optionals,' and error handling in Swift. They used Latent Dirichlet Allocation (LDA) to identify the topics from Stack Overflow questions and then cross-checked the findings by interviewing Swift developers. These are different from our research questions.

Zagalsky et al.~\citep{Zagalsky2016} have analyzed the R language using data from both Stack Overflow and R-help mailing list. They focused on the participation pattern of users in the two communities. They collected users' information in both sites and later mining on their activities (questions, answers). They tried to answer how communities create, share, and curates knowledge. Vasilescu et al.~\citep{Vasilescu2014} have compared popularity and user activity level between StackOverflow and R-help mailing list. They followed a similar approach to Zagalsky et al.~\citep{Zagalsky2016} by identifying active users in both communities. They have some interesting findings on the decreasing popularity of the mailing list and the influence of the reputation system in Stack Overflow. Their work is mainly focused on identifying the user behavior of these communities.

Vasilescu et al.~\cite{Vasilescu2013} have conducted a  study to find associations between software development and crowdsourced knowledge. They have found that the Stack Overflow activity rate correlates with the code changing activity in GitHub. One of their interesting findings is active GitHub committers ask fewer questions, but provided more answers than others.

In a study on developers' behavior, Xiong et al.~\cite{Xiong2017} have linked developers' activity across GitHub and Stack Overflow. They have shown that active issue committers are also active in asking questions. Moreover, for most developers, their contents on GitHub are similar to their questions and answers in the Stack Overflow.

Tausczik et al.~\citep{TausczikWC17} measured the effect of crowd size on Stack-Exchange question quality. They have found that among question audience size, contributor audience size, and topic audience size, contributor audience size has a higher effect on solution quality. They have classified the problems into three problems: error problems, how to problems, and conceptual problems. Error problems are very specific, and as a result, no matter how much the audience size is, 25\% problems are never solved. A large audience provides a diverse solution which is critical for how to problems. Conceptual problems are trickier and rarely solved with a small audience.

Srba et al.~\citep{Srba2016} had discussed the reason behind the increasing failure and churn rate of Stack Overflow. In their work, criticizing the existing automatic deletion and classification of posts, they introduced a new reputation system. They also suggested to follow answer oriented approach instead of the current asker-oriented approach. Instead of focusing on highly expert users, Stack Overflow should engage users of all levels.

%% file: Sections/Conclusion.tex
\section{Conclusion}
\label{sec:conclusion}
In this study, we have analyzed the reflection of the growth of new languages on Stack Overflow, i.e., how the activity pattern of Stack Overflow users changes along with the growth of the resources of the language and the expected time of availability of adequate resources. In the early stages of new programming languages, documentation is not very rich, and it is likely to be enriched with time. We have found that documentation of the language is one of the major topics about which developers talk about. The impact of the quality of documentation on the growth of new languages can be a new avenue for future work. We have also demonstrated a relationship between the growth of the three programming languages and developers' activity patterns using data from both Stack Overflow and GitHub. We have found that an active community can influence languages' growth and pinpointed the timeline after which language achieved enough resources for developers in QA sites. We believe our findings can help not only developers but also language owners and Stack Overflow to support the growth of new languages.